\documentclass[draft]{agujournal2019}
\usepackage{url} %this package should fix any errors with URLs in refs.
\usepackage{lineno}
\usepackage[inline]{trackchanges} %for better track changes. finalnew option will compile document with changes incorporated.
\usepackage{soul}
\usepackage{bm}
\usepackage{amssymb}
\usepackage{amsmath}

\draftfalse
\journalname{JGR: Planets}

\begin{document}

%%%%%%%%%%%%%%%%%%%%%%%%%%%%%%%%%%%%%%%%%%%%%%%
%  TITLE
%
% (A title should be specific, informative, and brief. Use
% abbreviations only if they are defined in the abstract. Titles that
% start with general keywords then specific terms are optimized in
% searches)
%
%%%%%%%%%%%%%%%%%%%%%%%%%%%%%%%%%%%%%%%%%%%%%%%

% Example: \title{This is a test title}

\title{Decameter-sized Earth Impactors -- II: A Bayesian Inference Approach to Meteoroid Ablation Modeling}
% \title{A Bayesian Inference Approach to Meteoroid Ablation Modeling}

%%%%%%%%%%%%%%%%%%%%%%%%%%%%%%%%%%%%%%%%%%%%%%%
%
%  AUTHORS AND AFFILIATIONS
%
%%%%%%%%%%%%%%%%%%%%%%%%%%%%%%%%%%%%%%%%%%%%%%%

% Authors are individuals who have significantly contributed to the
% research and preparation of the article. Group authors are allowed, if
% each author in the group is separately identified in an appendix.)

% List authors by first name or initial followed by last name and
% separated by commas. Use \affil{} to number affiliations, and
% \thanks{} for author notes.
% Additional author notes should be indicated with \thanks{} (for
% example, for current addresses).

% Example: \authors{A. B. Author\affil{1}\thanks{Current address, Antartica}, B. C. Author\affil{2,3}, and D. E.
% Author\affil{3,4}\thanks{Also funded by Monsanto.}}

\authors{Ian Chow\affil{1,2}, Peter G. Brown\affil{1,2}}

% \affiliation{1}{First Affiliation}
% \affiliation{2}{Second Affiliation}
% \affiliation{3}{Third Affiliation}
% \affiliation{4}{Fourth Affiliation}

\affiliation{1}{Department of Physics and Astronomy, University of Western Ontario, 1151 Richmond St, London, N6A 3K7, Ontario, Canada}
\affiliation{2}{Western Institute for Earth and Space Exploration, University of Western Ontario, Perth Drive, London, N6A 5B7, Ontario, Canada}
%(repeat as many times as is necessary)

% Corresponding author mailing address and e-mail address:

% (include name and email addresses of the corresponding author.  More
% than one corresponding author is allowed in this LaTeX file and for
% publication; but only one corresponding author is allowed in our
% editorial system.)

% Example: \correspondingauthor{First and Last Name}{email@address.edu}

\correspondingauthor{Ian Chow}{ichow9@uwo.ca}

%%%%%%%%%%%%%%%%%%%%%%%%%%%%%%%%%%%%%%%%%%%%%%%
% KEY POINTS
%%%%%%%%%%%%%%%%%%%%%%%%%%%%%%%%%%%%%%%%%%%%%%%
%  List up to three key points (at least one is required)
%  Key Points summarize the main points and conclusions of the article
%  Each must be 140 characters or fewer with no special characters or punctuation and must be complete sentences

% Example:
% \begin{keypoints}
% \item	List up to three key points (at least one is required)
% \item	Key Points summarize the main points and conclusions of the article
% \item	Each must be 140 characters or fewer with no special characters or punctuation and must be complete sentences
% \end{keypoints}

\begin{keypoints}
\item We develop a new method using dynamic nested sampling to estimate physical properties of meteoroids and asteroids from their light curves.
\item Using our new method, we identify three object classes in a structurally varied decameter impactor population, ranging from weak to strong.
\item Asteroids ranging from decimeter to decameter sizes both fragment in two distinct phases, but the latter do at much higher pressures.
\end{keypoints}

%%%%%%%%%%%%%%%%%%%%%%%%%%%%%%%%%%%%%%%%%%%%%%%
%
%  ABSTRACT and PLAIN LANGUAGE SUMMARY
%
% A good Abstract will begin with a short description of the problem
% being addressed, briefly describe the new data or analyses, then
% briefly states the main conclusion(s) and how they are supported and
% uncertainties.

% The Plain Language Summary should be written for a broad audience,
% including journalists and the science-interested public, that will not have 
% a background in your field.
%
% A Plain Language Summary is required in GRL, JGR: Planets, JGR: Biogeosciences,
% JGR: Oceans, G-Cubed, Reviews of Geophysics, and JAMES.
% see http://sharingscience.agu.org/creating-plain-language-summary/)
%
%%%%%%%%%%%%%%%%%%%%%%%%%%%%%%%%%%%%%%%%%%%%%%%

%% \begin{abstract} starts the second page

\begin{abstract}

Small asteroids and large meteoroids frequently impact the Earth, though their physical and material properties remain poorly understood.
When observed as fireballs in Earth's atmosphere, these properties can be inferred from their % inferred
ablation and fragmentation behavior.
The 2022 release of previously classified United States Government (USG) satellite sensor data has provided hundreds of new fireball light curves, allowing for more detailed analysis.
Here we present a new Bayesian inference method based on dynamic nested sampling that can robustly estimate these objects' physical parameters from their observed light curves, starting from relatively uninformative, flat priors. We validate our method against seven USG sensor-observed fireballs with independent ground-based observations and demonstrate that our results are consistent with previous estimates.
We then apply our technique to $13$ decameter-size Earth impactors to conduct the most detailed population-level study of their structure and material strength to date.
We identify three structurally distinct groups within the decameter impactors. The first group are primarily structurally homogeneous, weak objects which catastrophically disrupt below $\sim1.5$ MPa. The second group are heterogeneous objects which progressively fragment starting from $\sim1$ MPa typically up to $\sim3-8$ MPa. The third group are strong aggregates which remain mostly intact until $9-10$ MPa.
Our results also suggest that decameter-size asteroids fragment in two distinct phases: an initial phase at $\sim0.04-0.09$ MPa and a second at $\sim1-4$ MPa. 
While decimeter- to meter-size objects typically lose most of their mass in the initial phase, larger decameter-size objects instead lose most of their mass in the second phase.

\end{abstract}

\section*{Plain Language Summary}

% Small asteroids frequently impact the Earth, but their strength and internal structure still largely remain a mystery.
% Nevertheless, these impacts can cause significant ground damage and are the primary source of meteorites, underscoring the practical and scientific importance of understanding their physical makeup.

Small asteroids frequently impact the Earth, but their strength and internal structure are still poorly understood. However, asteroids can be observed burning up in the atmosphere as bright fireballs, which can provide information on their structure.
In 2022, data for hundreds of fireballs observed by United States Government (USG) satellite-based sensors were released to the public. Here we develop and validate a new statistical method to estimate the properties of impacting asteroids from their USG-recorded fireball light curves. We then apply this method to study the strength and structure of $13$ decameter-size (roughly $10$m-diameter) impactors, among the largest asteroids ever observed hitting the Earth.
We identify three main types of impactors. The first group are uniformly weak and quickly disintegrate below about $1.5$ MPa. The second group are made of material with a mix of strengths and gradually break up from about $1-8$ MPa. The third group are relatively strong and remain intact until about $9-10$ MPa. We also find large impactors break up in two stages, at about $0.04-0.09$ MPa and $1-4$ MPa. Smaller meteoroids typically lose most of their mass in the first stage, while the larger asteroids analyzed here instead do so in the second stage.

%%%%%%%%%%%%%%%%%%%%%%%%%%%%%%%%%%%%%%%%%%%%%%%
%
%  BODY TEXT
%
%%%%%%%%%%%%%%%%%%%%%%%%%%%%%%%%%%%%%%%%%%%%%%%

%%% Suggested section heads:
% \section{Introduction}
%
% The main text should start with an introduction. Except for short
% manuscripts (such as comments and replies), the text should be divided
% into sections, each with its own heading.

% Headings should be sentence fragments and do not begin with a
% lowercase letter or number. Examples of good headings are:

% \section{Materials and Methods}
% Here is text on Materials and Methods.
%
% \subsection{A descriptive heading about methods}
% More about Methods.
%
% \section{Data} (Or section title might be a descriptive heading about data)
%
% \section{Results} (Or section title might be a descriptive heading about the
% results)
%
% \section{Conclusions}

\section{Introduction}
Small asteroids ranging from $1-20$ meters in size impact the Earth $35-40$ times per year \cite{brown_flux_2002, bland_rate_2006, brown_500-kiloton_2013}, often appearing as spectacular fireballs in the Earth's atmosphere. 
% The largest of these objects can have kinetic energies equivalent to hundreds of kilotons of TNT, posing a hazard if they impact populated areas.
%...as shown by \citeA{jskilby}.
%...as shown by \citeA{lewin76}, \citeA{carson86}, \citeA{bartoldy02}, and \citeA{rinaldi03}.
%...has been shown \cite{jskilbye}.
%...has been shown \cite{lewin76,carson86,bartoldy02,rinaldi03}.
%... \cite <i.e.>[]{lewin76,carson86,bartoldy02,rinaldi03}.
%...has been shown by \cite <e.g.,>[and others]{lewin76}.
As most recovered meteorites originate from $1-20$ meter-size asteroids \cite{borovicka_are_2015}, this population in particular presents a unique opportunity to link data from fireball, telescopic and meteorite observations. The physical properties of these small asteroids have to date been poorly characterized at a population level, as they are often too small to be detected in space by telescopic surveys while also being comparatively rare as Earth impactors. However, the amount of data on the small asteroid population has grown significantly in recent years. In 2022, the U.S. Space Force publicly released decades of previously classified United States Government (USG) satellite sensor data on fireball events,
% (https://jpl.nasa.gov/news/us-space-force-releases-decades-of-bolide-data-to-nasa-for-planetary-defense-studies/)
including light curves of intensity over time. This new USG sensor data has increased the number of known impactors of meter size or larger to over one thousand, 
% ({\url{https://cneos.jpl.nasa.gov/fireballs/}), 
allowing for much deeper analysis at a population level. Since 2019, the Geostationary Lightning Mapper (GLM) instrument onboard the Geostationary Operational Environmental Satellite (GOES) has also recorded light curves for thousands of bright bolides \cite{mckinney_advancements_2025}, many of which are of meter size \cite{jenniskens_detection_2018}.

Recent telescopic surveys have also allowed a number of meter-size asteroids to be observed in space before impact. In cases where such pre-impact telescopic observations are available, combined analyses of such asteroids, as near-Earth objects (NEOs) in space, fireballs in Earth's atmosphere and in select instances as recovered meteorites have allowed them to be characterized in great detail \cite<e.g.>[]{jenniskens_impact_2009, jenniskens_impact_2021, geng_near-earth_2023, clark_preatmospheric_2023, spurny_atmospheric_2024, kareta_telescope--fireball_2024, gianotto_fall_2025, ingebretsen_apache_2025, egal_catastrophic_2025}.
Upcoming surveys such as NEO Surveyor \cite{mainzer_near-earth_2023} and the Vera C. Rubin Observatory's Legacy Survey of Space and Time \cite<LSST;>[]{ivezic_lsst_2019} are expected to discover over $100,000$ NEOs each \cite{mainzer_near-earth_2023, kurlander_predictions_2025}, with LSST expected to discover $\sim1-2$ imminent Earth impactors per year (Chow et al., 2026, submitted).

The rapidly growing amount of data on these impactors thus presents a unique opportunity to conduct population-level analysis of their properties for the first time. 
Modern studies using light curve data to forward model and invert for the physical and material properties of specific meteorite-producing fireballs have typically proceeded by generating synthetic light curves by simulating the object's atmospheric ablation and fragmentation and then manually fitting the simulated light curve to observations by adjusting various model parameters. \cite<e.g.>[]{borovicka_kosice_2013, wheeler_fragment-cloud_2017, mcmullan_uncertainty_2019, mcfadden_comparison_2024}.
However, this manual fitting approach is labor-intensive, subject to parameter degeneracy and does not characterize uncertainties in these objects' inferred properties. Previous attempts to develop automated approaches for ablation modelling using genetic algorithms \cite{tarano_inference_2019, henych_semi-automatic_2023} have seen only limited success for a small number of fireballs, require an initial manual solution to be found first and often need to be run numerous times per fireball to obtain a good fit to observations.
Motivated by this recent release of USG sensor data, we therefore develop a novel automated Bayesian inference approach to meteoroid ablation modeling that uses dynamic nested sampling \cite{skilling_nested_2004, skilling_nested_2006, higson_dynamic_2019} 
in conjunction with the semi-empirical fragmentation model of \citeA{borovicka_kosice_2013} 
to probabilistically estimate the physical and material properties of these fireballs solely from their light curves. 
Compared to previous manual fitting or genetic algorithm methods, our approach is capable of producing accurate fits from a single run under relatively uninformative flat priors and allows for robust uncertainty estimation in the physical properties of meteoroids.
We validate our method against seven fireball events with USG light curves and for which detailed entry modeling has previously been conducted using independent ground-based observations, and demonstrate that our results are consistent with previous estimates.
We then apply our method to $13$ decameter-size Earth impactors previously identified in \citeA{chow_decameter-sized_2025}, where for all but one case (the 15 February 2013 Chelyabinsk bolide) only USG light curve data is available, to characterize the physical and material properties of these objects at a population level for the first time.

This paper is structured as follows: Section \ref{sec:data} introduces the USG sensor data used. Section \ref{sec:model} describes our light curve fitting procedure using the semi-empirical model and the nested sampling approach for parameter inference.
In Section \ref{sec:ground-based}, we validate our nested sampling method against seven USG sensor-recorded fireballs for which entry modeling has been previously conducted using independent ground-based observations, and demonstrate that our results are generally consistent with previous estimates. We then apply our method in Section \ref{sec:decameter_impactors} to infer physical and material properties of $13$ decameter-size impactors previously identified in \citeA{chow_decameter-sized_2025} and connect them to their dynamical origins. We discuss our results in Section \ref{sec:discussion}. Finally, we summarize our conclusions and discuss future applications of the nested sampling method in Section \ref{sec:conclusions}.

% We test our Bayesian inference method against several USG-recorded fireballs with previously fit solutions based on independent ground-based observations, and demonstrate that the results from our method compare favourably to the manual solutions. We then apply this method to the $14$ decameter-size impactors previously identified in \citeA{chow_decameter-sized_2025}.

% providing a wealth of information about their physical and material properties

% Numerous studies have used light curve data of these objects' impacts to analyze their ablation in the atmosphere, often in conjunction with various semi-analytic models for atmospheric fragmentation \cite<e.g.>[]{chyba_1908_1993, hills_fragmentation_1993, artemieva_motion_2001, avramenko_simulation_2014, wheeler_fragment-cloud_2017, wheeler_atmospheric_2018} \noteic{could shorten this if needed...}, to uncover a wealth of understanding about their physical and material properties \cite<e.g.>[]{borovicka_kosice_2013, borovicka_two_2020, mcfadden_comparison_2024} \noteic{these are off the top of my head, add more?}. 

% Indeed, a number of semi-analytic models for asteroid fragmentation \cite<e.g.>[]{chyba_1908_1993, hills_fragmentation_1993, artemieva_motion_2001, avramenko_simulation_2014, wheeler_fragment-cloud_2017, wheeler_atmospheric_2018} have been developed to model 

\section{Data}\label{sec:data}

NASA's Center for Near Earth Object Studies (CNEOS) Fireball and Bolide Database reports fireball events detected by satellite-based USG sensors, including light curves of fireball intensity as a function of time. From the only published information in the literature, the USG sensors observe in the optical passband from approximately $400-1100$ nm at a sampling rate of $333$ Hz and the total optical energy is converted from the sensor passband under the assumption that the overall spectral energy distribution of each fireball resembles a $6000$ K blackbody \cite{tagliaferri_analysis_1995}. The total radiated energy and computed impact energy are recorded by CNEOS as well, and are generally consistent with energies determined through ground-based optical observations \cite{devillepoix_observation_2019}, infrasound measurements \cite{gi_refinement_2017} and common fireball detections by the GLM \cite{wisniewski_determining_2024}. The heights of peak brightness also tend to be well estimated, with comparisons to ground-based observations suggesting errors are typically of order $2-3$ km or less in most cases. Radiant information and speed are less accurately measured. This is discussed extensively in \citeA{chow_decameter-sized_2025}.
We note that while fireball dynamics is a powerful tool for model constraints for smaller objects \cite{ceplecha_atmospheric_1993}, for the very large fireballs discussed here there is usually little to no observable deceleration prior to major fragmentation (and often little afterward) so the main information about the body is contained in the light curve. 
The light curves used for modeling in this work are primarily taken from the CNEOS database with the exception of the Chelyabinsk bolide, for which we use the light curve of \citeA{brown_500-kiloton_2013} as the USG sensor light curve is of poor quality.
% In some cases we specify when the data has been supplemented by other sources.
% if the USG sensor light curve is unavailable or of poor quality. 

\section{Methods}\label{sec:model}

In this section we describe our procedure for modeling the observed USG light curves. We use as our physical ablation model the semi-empirical model of meteoroid ablation and fragmentation developed by \citeA{borovicka_kosice_2013} and employ it in conjunction with dynamic nested sampling to estimate the posterior distribution of the model parameters. 

\subsection{Nested Sampling}

Modern astronomy often requires inferring characteristics of physical models based on observational data. In recent decades, the rise of large astronomical datasets and concurrent advances in computational power for processing them have resulted in a paradigm shift for astrophysical inference problems, from traditional frequentist methods such as maximum likelihood estimation \cite{fisher_mathematical_1922} to Bayesian methods which aim to estimate the posterior distribution of a set of model parameters given some prior knowledge and observed data.

Under a Bayesian framework, the posterior distribution of a vector of parameters $\bm\Theta$ for a given model $M$ conditioned on observed data $D$ is given by Bayes' theorem
\begin{equation}
    P\left(\bm\Theta|D, M\right) = \frac{P\left(D|\bm\Theta, M\right)\,P\left(\bm\Theta|M\right)}{P\left(D|M\right)} \label{eqn:bayes}
\end{equation}
where $P\left(D|\bm\Theta, M\right) = \mathcal{L}\left(\bm\Theta\right)$ is the likelihood function, $P\left(\bm\Theta|M\right) = \pi\left(\bm\Theta\right)$ is the prior and $P\left(D|M\right) = \mathcal{Z}$ is the Bayesian evidence (or marginal likelihood), a term which quantifies how well the data supports the specific model. For complicated, high-dimensional models, such as those used for simulating meteoroid ablation, the posterior is often analytically intractable, necessitating the use of numerical methods to estimate the posterior using a finite number of weighted points.

One such method is nested sampling \cite{skilling_nested_2004, skilling_nested_2006}, which estimates both the Bayesian evidence as well as the posterior distribution. Nested sampling works by dividing the posterior distribution into nested ``slices" with contours of constant likelihood and sequentially sampling from each slice before combining the samples (with appropriate weights) to estimate the overall posterior. Nested sampling is effective for high-dimensional problems with multimodal posteriors, where other sampling techniques popular for astrophysical inference such as Markov Chain Monte Carlo \cite<MCMC;>[]{goodman_ensemble_2010, foreman-mackey_emcee_2013} often perform poorly. This is particularly salient for meteoroid ablation modelling, where previous studies have typically noted that the manually determined solution is not unique and can often be reproduced by many different physically reasonable fragment sequences or parameters \cite<e.g.>[]{borovicka_kosice_2013, borovicka_january_2017, borovicka_maribo_2019, henych_semi-automatic_2023}.

Dynamic nested sampling \cite{higson_dynamic_2019} is a modification of nested sampling that adaptively allocates live points during the sampling process to favor sampling particular regions of parameter space. By increasing the number of points in regions where the majority of the posterior ``mass" is located and reducing the number of points elsewhere, dynamic nested sampling can produce a more precise estimate for the posterior distribution at the cost of a less precise estimate for the Bayesian evidence. In this work we are primarily interested in the posterior distribution of the model parameters rather than the evidence and as such adopt the dynamic nested sampling paradigm for inference throughout.

\subsection{Meteoroid Ablation Model}

We use the semi-empirical model of meteoroid ablation and fragmentation first introduced by \citeA{borovicka_kosice_2013} which is grounded in earlier work that focused on fireball gross fragmentation dynamics and ablation \cite{ceplecha_atmospheric_1993} and uses the modern synthesis for intrinsic/apparent ablation and luminous efficiency described in \citeA{ceplecha_fragmentation_2005}. As the name implies, it is an empirically adapted version of single-body ablation informed by precise observations of fireballs, with particular focus on describing fireball fragmentation as a physical process with model outputs tuned to be directly comparable to direct measurements of fireballs. 
% to model the Ko\v{s}ice fireball 
In the model, the positions of fragmentation points can be adjusted as required, with each fragmentation producing either monolithic single-body fragments, dust, or eroding fragments which are major fragments that gradually release dust over a period of time after ejection from the main body. The individual light curves produced by the ablation of each fragment or dust particle are then computed and summed to form the total light curve for the object. Atmospheric densities are taken from the empirical NRLMSISE-00 model of \citeA{picone_nrlmsise-00_2002}, evaluated at the time and location of each fireball. More details of the semi-empirical model are given in \citeA{popova_modelling_2019} and \citeA{borovicka_two_2020}. All light curve modeling in this paper is conducted using the \texttt{MetSim} software \cite{vida_direct_2023}.

Our model fits involve finding suitable physical parameters which reproduce the observed light curve as a function of height. In our fitting process, we first manually select one or more approximate fragmentation heights $H_j$ for each impactor based on the shape of the USG lightcurve. For the validation cases, we associate these USG-based estimates with the major fragmentation points identified in previous studies. To simplify the model, we assume that every fragmentation produces an eroding fragment, as the release of single-body fragments or dust can be approximated as end member cases by using a very low or very high erosion coefficient, respectively. Each fragment is then parameterized by its fragmentation height $h$, mass as a percentage of the initial meteoroid mass $m_\mathrm{frag}$, erosion coefficient $\eta$, grain minimum mass $m_\mathrm{lower}$ and grain maximum mass $m_\mathrm{upper}$. The initial meteoroid mass $m_\mathrm{init}$ is also included as a free parameter in our model. We adopt a fixed initial bulk density of $1500$ kg m$^{-3}$ based on other estimates of NEO bulk densities \cite<see Section $2$ of>[and references therein]{chow_decameter-sized_2025} and a fixed grain density of $3500$ kg m$^{-3}$ appropriate for chondritic meteorites \cite{flynn_physical_2018}. 
While we recognize there is some variance in the bulk density of these impactors, it is difficult to quantify. However, as discussed in Section \ref{sec:uncertainties_limitations}, we empirically find that this uncertainty in the bulk density does not significantly affect the results of our analysis.

% we use $1500$ kg m$^{-3}$ for consistency with 

For each impactor, we use a fixed initial velocity $v$ based on the USG-reported entry velocity and a fixed luminous efficiency $\tau$ based on the optical energy-$\tau$ function given in \citeA{brown_flux_2002}. Finally, we also use a fixed shape parameter of $\Gamma A = 0.8$ following \citeA{borovicka_two_2020}, a mass distribution index of $s = 2.5$ and an ablation coefficient of $0.005$ kg MJ$^{-1}$ across all fragments and all fireballs as found to be representative for chondritic meteoroids \cite{ceplecha_fragmentation_2005, borovicka_two_2020}.

The free parameters determining the light curve data for a model with $k$ fragmentation points are thus $\bm{\Theta} = \{m_\mathrm{init}\} \cup \{m_{\mathrm{frag}_j}, \eta_j, m_{\mathrm{lower}_j}, m_{\mathrm{upper}_j}, h_j\}_{j=1}^{k}$. Assuming Gaussian uncertainty in the observations, the log-likelihood $\ln\mathcal{L}$ of a given set of parameters $\bm{\Theta}$ is then written as

\begin{equation}
    \ln\mathcal{L}\left(\bm\Theta\right) = -\sum_{i=1}^{N}\left(\frac{\left(L_i - \bar{L}_i\left(\bm{\Theta}\right)\right)^2}{2\sigma_i^2} + \ln\sqrt{2\pi\sigma_i^2}\right),\label{eqn:logL}
\end{equation}

where $L_i$ denotes the $i$th observed luminosity sample, $\sigma_i^2$ denotes the corresponding measurement uncertainty, and $\bar{L}_i\left(\bm{\Theta}\right)$ denotes the simulated luminosity from the ablation model at the height of observation. 

The posterior distribution of the free parameters $\bm\Theta$ is then estimated using dynamic nested sampling. In our model, the fragments' grain minima $m_{\mathrm{lower}_j}$, grain maxima $m_{\mathrm{upper}_j}$, and the initial meteoroid mass $m_{\mathrm{init}}$ are parameterized by their logarithms $\log_{10}m_{\mathrm{lower}_j}$, $\log_{10}m_{\mathrm{upper}_j}$ and $\log_{10}m_{\mathrm{init}}$ respectively. We adopt uniform priors for all model parameters over their respective intervals, summarized in Table \ref{tab:model_params}. Our priors are relatively uninformative and are chosen to ensure that the full range of physically reasonable parameter values are contained.
Note that the lower bound for the grain maximum is set so it cannot be smaller than the grain minimum. The bounds on the fragmentation heights $h_j$ are set based on the manually selected fragmentation heights $H_j$ as the smaller of $5$ kilometers or half the vertical distance to the previous or next manually selected fragmentation point (if applicable), to ensure that fragmentation heights are always in strictly descending order.

% For the $j$th fragment, we adopt uniform priors on the mass percentage $m_{\mathrm{frag}_j}$ over the interval $\left[0, 100\right]$ and on the erosion coefficient $\eta_j$ over the interval $\left[0.01, 10\right]$ s$^2$ km$^{-2}$. 
% We adopt uniform priors for the log of the grain minimum, $\log_{10}m_{\mathrm{lower}_j}$, over the interval [$-5$, $2$] $\log_{10}\left(\mathrm{kg}\right)$ and the log of the grain maximum, $\log_{10}m_{\mathrm{upper}_j}$, over the interval $\left[\max\left(\log_{10}m_{\mathrm{lower}_j}, -4\right), 3\right]$ $\log_{10}\left(\mathrm{kg}\right)$. The latter constraint is imposed as the grain maximum cannot be smaller than the grain minimum. 
% The prior on the fragmentation height, $h_j$, is uniform between $\left[\max\left(H_j - 5\,\mathrm{km}, \frac{H_{j}-H_{j-1}}{2}\right), \min\left(H_j + 5\,\mathrm{km}, \frac{H_{j+1}-H_{j}}{2}\right)\right]$ where $H_j$ is the manually chosen $j$th fragmentation height (i.e. the smaller of 5 kilometers or half the vertical distance to the previous/next fragmentation point). This is done to ensure that the fragmentation heights are always in strictly descending order.
% Finally, the prior on the log of the initial mass, $\log_{10}m_{\mathrm{init}}$, is uniform between $\left[2, 8\right]$ $\log_{10}\left(\mathrm{kg}\right)$. 

\begin{table*}[t]
    \begin{center}
    % \caption{Nested Sampling Model Parameter Uniform Prior Bounds}
    \begin{tabular}{|l|c c|}
    %%%% FIXED PARAMETERS
    % \hline
    % \textbf{Fixed Parameter} & \multicolumn{2}{c|}{}} \\
    %%%% FREE PARAMETERS
    \hline
    \textbf{Parameter} & \textbf{Lower Bound} & \textbf{Upper Bound} \\
    \hline
    $\log_{10}m_{\mathrm{init}}$ ($\log_{10}\mathrm{kg}$) & $2$ & $8$ \\
    \hline
    \textbf{For $j$th fragment:} & & \\
    \hline
    $m_{\mathrm{frag}_j}$ (\% of $m_{\mathrm{init}}$) & $0$ & $100$ \\
    $\eta_j$ (s$^2$ km$^{-2}$) & $0.01$ & $10$ \\
    $\log_{10}m_{\mathrm{lower}_j}$ ($\log_{10}\mathrm{kg}$) & $-5$ & $2$ \\
    $\log_{10}m_{\mathrm{upper}_j}$ ($\log_{10}\mathrm{kg}$) & $\max\left(\log_{10}m_{\mathrm{lower}_j}, -4\right)$ & $3$ \\
    $h_j$ (km) & $\max\left(H_j - 5, \frac{H_{j}+H_{j-1}}{2}\right)$ & $\min\left(H_j + 5, \frac{H_{j+1}+H_{j}}{2}\right)$ \\ \hline
    \end{tabular}
    \end{center}
\caption{Prior bounds for all free model parameters used for nested sampling. We adopt uniform priors over all model parameters with the specified lower and upper bounds. The lower bound for the fragment grain maximum $m_{\mathrm{upper}_j}$ is imposed based on the chosen grain minimum $m_{\mathrm{lower}_j}$ (as the maximum must be greater than the minimum), and the bounds on the fragmentation heights $h_j$ are set based on $H_j$, the \textit{manually} selected fragmentation heights, to ensure that they remain in strictly descending order.}
    \label{tab:model_params}
\end{table*}

The USG sensor data does not include observational uncertainties. We therefore estimate a fixed measurement uncertainty of $\sigma_i = 1.2\times 10^{10}$ W ster$^{-1}$ for all USG observations in this paper, based on the detection limit of USG sensors at magnitude $\sim-16.5$ assuming a bolometric power of $3030$ W at zero magnitude. 
% The intrinsic scatter can ideally be estimated with the other parameters by introducing a scatter term $\sigma_{\mathrm{scat}}^2$ to the uncertainty in the log-likelihood and allowing it to vary as a free parameter in the model. However, our efforts to incorporate scatter in this manner into the nested sampling model for this work were unsuccessful as we discuss in Section \ref{sec:discussion} and as such we use the log-likelihood formalism of Equation \ref{eqn:logL} throughout.
For each event, the dynamic nested sampling is then conducted with the \texttt{dynesty} \cite{speagle_dynesty_2020, sergey_koposov_joshspeagledynesty_2024} software package using $1000$ live points and the random-slice sampling strategy \cite{neal_slice_2003, handley_polychord_2015, handley_polychord_2015-1}, with the stopping condition and other hyperparameters set to default values.

% The resulting RV fit is shown in Figure 1. The top panel shows the 1σ,2σ, and 3σ uncertainties of our fit RV curve together with the observed data. We obtain best-fit instrumental jitter values of 1.41, 0.694, and 3.00 m s−1 for the HARPS1, HARPS2, and HIRES telescopes, respectively. The maximum log-likelihood solution and the 5%, 50%, and 95% quantiles of the 1D posterior distributions obtained from MCMC sampling are reported in Table

% Our method using the \texttt{dynesty} \cite{speagle_dynesty_2020, sergey_koposov_joshspeagledynesty_2024} dynamic nested sampling \cite{skilling_nested_2004, skilling_nested_2006, higson_dynamic_2019} software in conjunction with the \texttt{MetSim} software \cite{vida_direct_2023} to characterize uncertainties in the physical properties of these objects for the first time.

\section{Validation Against Ground-Based Observations}\label{sec:ground-based}

Here we validate our method against all seven fireballs (listed in chronological order) for which detailed light curve modelling has previously been conducted using independent ground-based observations, to compare the range of the parameter values derived from our nested sampling-based method to estimates of previous works. For each event, the maximum log-likelihood solution and the $5\%$, $50\%$ (median) and $95\%$ quantiles of the $1$D posterior distributions for several relevant physical quantities of the object derived from nested sampling are recorded. These quantities are the initial object mass $m_\mathrm{init}$, the dynamic pressure at the main fragmentation point $P_{\mathrm{main}}$, the mass released at the main fragmentation point $m_\mathrm{main}$, the peak dynamic pressure experienced $P_{\mathrm{peak}}$, and the mass remaining at peak dynamic pressure $m_{\mathrm{peak}}$. 
Note that the choice of luminous efficiency model is generally the main driver scaling differences in initial mass. Here we adopt the empirical integral luminous efficiency model from \citeA{brown_flux_2002} as it was calibrated using USG-recorded events from USG sensors. This model sets luminous efficiency to be independent of mass and velocity, depending only on the initial optical energy of the fireball. The more detailed luminous efficiency model of \citeA{borovicka_two_2020}, which is part of the semi-empirical model framework, has much better validation at small masses and offers more precise overall measurements (including terms for velocity and mass) but for much smaller fireballs than considered here. We summarize our results in Table \ref{tab:validation} and discuss each event individually in the following sections.

\begin{table*} 
\makebox[1 \textwidth][r]{       %centering table
\resizebox{1.2 \textwidth}{!}{   %resize table
\begin{tabular}{|l|c c c|c c c|c c c|c c c|c c c|}
\hline
\textbf{Event} 
& \multicolumn{3}{c|}{{$\bm{m_\mathrm{init}}$ (kg)}}
& \multicolumn{3}{c|}{{$\bm{P_\mathrm{main}}$ (MPa)}}
& \multicolumn{3}{c|}{{$\bm{m_\mathrm{main}}$ (\% of $m_\mathrm{init}$)}}
& \multicolumn{3}{c|}{{$\bm{P_\mathrm{peak}}$ (MPa)}}
& \multicolumn{3}{c|}{{$\bm{m_\mathrm{peak}}$ (\% of $m_\mathrm{init}$)}} \\
&
$5\%$ & $50\%$ & $95\%$ &
$5\%$ & $50\%$ & $95\%$ &
$5\%$ & $50\%$ & $95\%$ &
$5\%$ & $50\%$ & $95\%$ &
$5\%$ & $50\%$ & $95\%$
\\
\hline
\textbf{Tagish Lake} & & & & & & & & & & & & & & & \\
(1) & \multicolumn{3}{c|}{$33000-100000$} & \multicolumn{3}{c|}{$0.7$} & \multicolumn{3}{c|}{$46$} & \multicolumn{3}{c|}{$1.2$} & \multicolumn{3}{c|}{$3.9$} \\
This work max. $\log\mathcal{L}$ & \multicolumn{3}{c|}{$95000$} & \multicolumn{3}{c|}{$1.0$} & \multicolumn{3}{c|}{$41$} & \multicolumn{3}{c|}{$3.1$} & \multicolumn{3}{c|}{$2.1$} \\
This work quantiles & $95000$ & $98000$ & $100000$ & $0.9$ & $1.0$ & $1.0$ & $41$ & $47$ & $53$ & $3.0$ & $3.1$ & $5.2$ & $1.9$ & $2.5$ & $3.6$ \\
% cut 45k out of 79k
\hline
\textbf{Mor\'{a}vka} & & & & & & & & & & & & & & & \\
(2) & \multicolumn{3}{c|}{$230-11000$} & \multicolumn{3}{c|}{$4.3$} & \multicolumn{3}{c|}{$48$} & \multicolumn{3}{c|}{$5.0$} & \multicolumn{3}{c|}{$18$} \\
This work max. $\log\mathcal{L}$ & \multicolumn{3}{c|}{$3000$} & \multicolumn{3}{c|}{$4.0$} & \multicolumn{3}{c|}{$34$} & \multicolumn{3}{c|}{$8.9$} & \multicolumn{3}{c|}{$26$} \\
This work quantiles & $2100$ & $3000$ & $5100$ & $2.6$ & $3.9$ & $4.5$ & $19$ & $38$ & $80$ & $5.5$ & $8.7$ & $11$ & $11$ & $26$ & $39$ \\
% cut 2.5k out of 41k
\hline
\textbf{Park Forest} & & & & & & & & & & & & & & & \\
(3) & \multicolumn{3}{c|}{$7000-21000$} & \multicolumn{3}{c|}{$5.38-6.99$} & \multicolumn{3}{c|}{$63$} & \multicolumn{3}{c|}{$14.0$} & \multicolumn{3}{c|}{$0.84$} \\
This work max. $\log\mathcal{L}$ & \multicolumn{3}{c|}{$13000$} & \multicolumn{3}{c|}{$6.26$} & \multicolumn{3}{c|}{$87$} & \multicolumn{3}{c|}{$8.73$} & \multicolumn{3}{c|}{$1.0$} \\
This work quantiles & $13000$ & $13000$ & $15000$ & $6.18$ & $6.26$ & $6.35$ & $79$ & $87$ & $88$ & $8.00$ & $8.67$ & $10.6$ & $0.92$ & $3.8$ & $11$ \\
% cut 20k out of 50k
\hline
\textbf{Ko\v{s}ice} & & & & & & & & & & & & & & & \\
% 90.0 or 86.6 % given initial mass?
(4) & \multicolumn{3}{c|}{$1200-11000$} & \multicolumn{3}{c|}{$0.97$} & \multicolumn{3}{c|}{$91$} & \multicolumn{3}{c|}{$5.9$} & \multicolumn{3}{c|}{$0.52$} \\
This work max. $\log\mathcal{L}$ & \multicolumn{3}{c|}{$17000$} & \multicolumn{3}{c|}{$0.91$} & \multicolumn{3}{c|}{$96$} & \multicolumn{3}{c|}{$6.7$} & \multicolumn{3}{c|}{$1.4$} \\
This work quantiles & $16000$ & $17000$ & $22000$ & $0.85$ & $0.91$ & $1.00$ & $79$ & $97$ & $99$ & $3.5$ & $5.9$ & $15$ & $0.086$ & $1.2$ & $10$ \\
% (4) & & $1167$--$10500$ & & & $0.97$ & & & $57$ & & & $5.9$ & & & $0.5$ & \\
% This work max. $\log\mathcal{L}$ & \multicolumn{3}{c|}{$17504$} & \multicolumn{3}{c|}{$0.88$} & \multicolumn{3}{c|}{$96.6$} & \multicolumn{3}{c|}{$7.44$} & \multicolumn{3}{c|}{$1.2$} \\
% This work quantiles & $16753$ & $17721$ & $22558$ & $0.82$ & $0.88$ & $0.95$ & $76.9$ & $95.4$ & $98.7$ & $3.4$ & $8.2$ & $15.5$ & $1.2$ & $2.9$ & $14.0$ \\
% cut 15k out of 45k
\hline
\textbf{Romania} & & & & & & & & & & & & & & & \\
(5) & \multicolumn{3}{c|}{$3000-6800$} & \multicolumn{3}{c|}{$1.0$} & \multicolumn{3}{c|}{$90$} & \multicolumn{3}{c|}{$3.0$} & \multicolumn{3}{c|}{$\sim0$} \\
This work max. $\log\mathcal{L}$ & \multicolumn{3}{c|}{$13000$} & \multicolumn{3}{c|}{$1.0$} & \multicolumn{3}{c|}{$84$} & \multicolumn{3}{c|}{$25$} & \multicolumn{3}{c|}{$0.80$} \\
% (5) & & $3000$--$6750$ & & & $1.0$ & & & $90$ & & & $3.0$ & & & $\sim0$ & \\
This work quantiles & $6100$ & $12000$ & $15000$ & $0.9$ & $1.0$ & $1.1$ & $60$ & $79$ & $86$ & $20$ & $25$ & $27$ & $0.56$ & $0.87$ & $1.7$ \\
% cut 30k out of 97k
\hline
\textbf{Sari\c{c}i\c{c}ek} & & & & & & & & & & & & & & & \\
(6) & \multicolumn{3}{c|}{$780-2700$} & \multicolumn{3}{c|}{$1.00$} & \multicolumn{3}{c|}{not reported} & \multicolumn{3}{c|}{$3.99$} & \multicolumn{3}{c|}{$1.6$} \\
This work max. $\log\mathcal{L}$ & \multicolumn{3}{c|}{$2300$} & \multicolumn{3}{c|}{$1.42$} & \multicolumn{3}{c|}{$53$} & \multicolumn{3}{c|}{$4.93$} & \multicolumn{3}{c|}{$1.9$} \\
This work quantiles & $2300$ & $2500$ & $2900$ & $1.29$ & $1.42$ & $1.49$ & $44$ & $55$ & $91$ & $4.11$ & $6.11$ & $9.73$ & $1.6$ & $2.9$ & $7.7$ \\
% (6) & & $2510$--$3770$\footnote{Computed from given diameter assuming a bulk density of $1500$kg m$^{-3}$} & & & $1.0$ & & & not given & & & $4.0$ & & & not given & \\
% This work max. $\log\mathcal{L}$ & \multicolumn{3}{c|}{$2670$} & \multicolumn{3}{c|}{$1.39$} & \multicolumn{3}{c|}{$51.6$} & \multicolumn{3}{c|}{$3.22$} & \multicolumn{3}{c|}{$4.2$} \\
% This work quantiles & $2378$ & $2590$ & $5687$ & $1.25$ & $1.35$ & $1.47$ & $45.8$ & $60.9$ & $88.5$ & $2.82$ & $3.66$ & $11.00$ & $2.1$ & $3.3$ & $9.2$ \\
% cut 10k out of ~87k
\hline
\textbf{Flensburg} & & & & & & & & & & & & & & & \\
(7) & \multicolumn{3}{c|}{$10000-22000$} & \multicolumn{3}{c|}{$0.7$} & \multicolumn{3}{c|}{$68$} & \multicolumn{3}{c|}{$2.0$} & \multicolumn{3}{c|}{$3$} \\
This work max. $\log\mathcal{L}$ & \multicolumn{3}{c|}{$18000$} & \multicolumn{3}{c|}{$0.7$} & \multicolumn{3}{c|}{$60$} & \multicolumn{3}{c|}{$8.2$} & \multicolumn{3}{c|}{$10$} \\
This work quantiles & $12000$ & $14000$ & $19000$ & $0.5$ & $0.7$ & $0.7$ & $54$ & $77$ & $97$ & $2.7$ & $7.2$ & $8.8$ & $3$ & $10$ & $10$ \\
% cut 12k of ~40k
\hline
\end{tabular}
}
}

% \caption{Comparison of Maximum Log-Likelihood and Nested Sampling Posterior Quantiles for Derived Physical Parameters of Validation Events to Previous Estimates}
\caption{
The maximum log-likelihood fit and nested sampling posterior quantiles for physical parameters of seven USG sensor-recorded fireball events compared to previous parameter estimates derived using independent ground-based observations, with corresponding references provided. The parameters are the initial mass $m_\mathrm{init}$, dynamic pressure at main fragmentation $P_\mathrm{main}$, mass released at main fragmentation $m_\mathrm{main}$, peak dynamic pressure $P_\mathrm{peak}$ and mass remaining at peak dynamic pressure $m_\mathrm{peak}$. These seven events represent all fireballs both observed by USG sensors and for which detailed light curve modeling has previously been conducted using the independent observations. Computed results are rounded to the same number of significant digits reported by previous studies.}
\textbf{References}: (1) \citeA{brown_entry_2002} and \citeA{hildebrand_fall_2006}; (2) \citeA{neder_radionuclide_2001}, \citeA{borovicka_moravka_2003}, \citeA{borovicka_moravka_2003-1} and \citeA{borovicka_moravka_2003-2}; (3) \citeA{brown_orbit_2004}; (4) \citeA{borovicka_kosice_2013}; (5) \citeA{borovicka_january_2017}; (6) \citeA{unsalan_saricicek_2019}; (7) \citeA{borovicka_trajectory_2021}.% , J. Borovi\v{c}ka, personal communication (2025)  
\label{tab:validation}
\end{table*}

\subsection{Tagish Lake}\label{sec:tagish_lake}

The Tagish Lake meteorite fell in Canada on 18 January 2000. It was accompanied by a bright fireball detected by USG sensors, seismic and infrasonic stations and recorded by eyewitnesses on video \cite{brown_entry_2002, hildebrand_fall_2006}. The pre-impact mass of the object was estimated to be between $33-100$ tons by \citeA{brown_entry_2002} using several different methods. Modeling the USG sensor light curve with the gross fragmentation model of \citeA{ceplecha_atmospheric_1993} (a precursor to the semi-empirical model), \citeA{brown_entry_2002} placed the first major burst point at a height of $37$ km under a dynamic pressure of $0.7$ MPa. The meteoroid reached its peak dynamic pressure of $1.2$ MPa at a height of $\sim30.8$ km. While fragment masses are not reported, \citeA{brown_entry_2002} also estimated that $\sim26$ tons of the original $56$-ton mass remained at the $37$-km burst point (which we take as the mass released at the main fragmentation) and that $\sim2.2$ tons remained at peak dynamic pressure.

\begin{figure*}
\centering
\includegraphics[width=1.\linewidth]{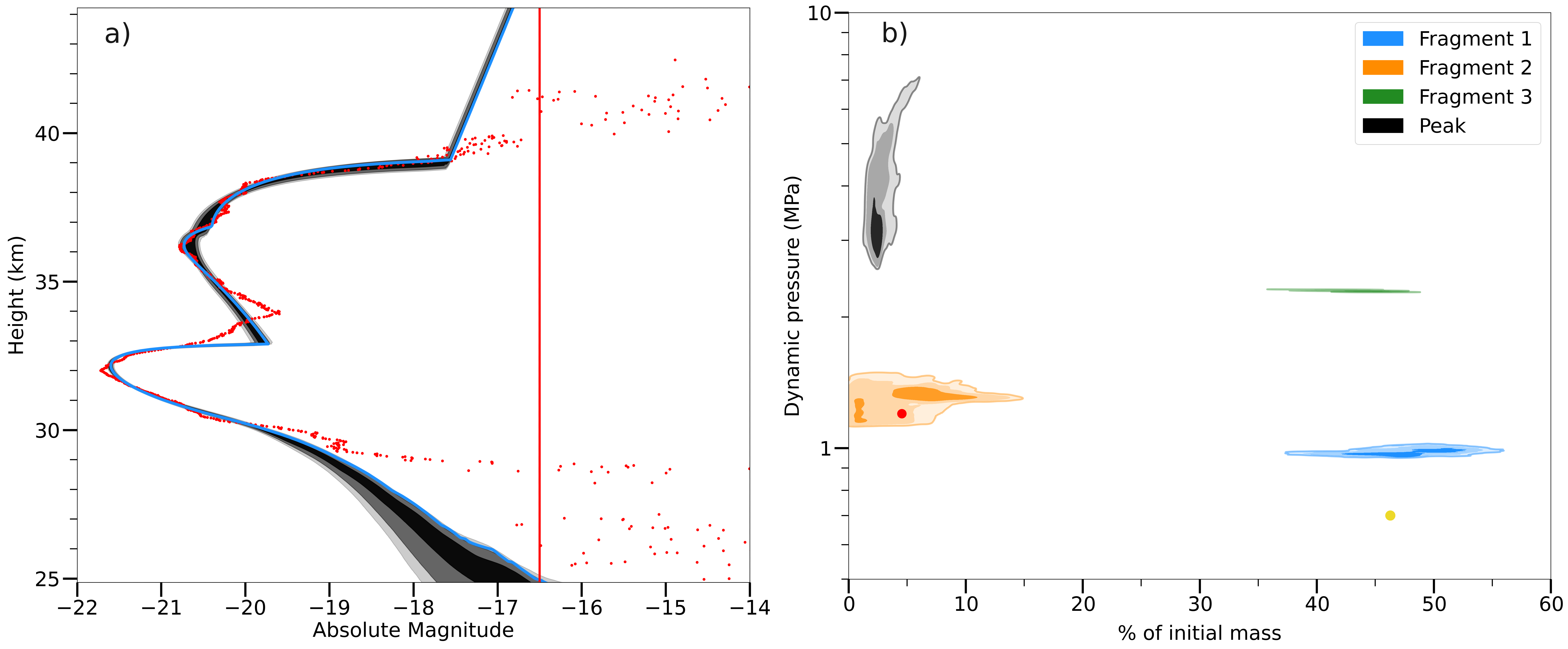}
    \caption{
    \textbf{a)}: The model light curve of intensity versus height plotted over the USG sensor-recorded fireball light curve (red dots) for the 18 January 2000 Tagish Lake meteorite-producing fireball. 
    Here the black shaded regions illustrate the $1\sigma$, $2\sigma$ and $3\sigma$ distributions for the light curve solutions derived from nested sampling.
    % Here $1\sigma$, $2\sigma$ and $3\sigma$ uncertainties in the model fit are illustrated by the black shaded regions. 
    The maximum log-likelihood solution obtained by nested sampling is plotted as the blue line. The intensity is given in units of absolute stellar magnitudes assuming a bolometric power of $3030$ W at zero magnitude. The detection limit of USG sensors at absolute magnitude $-16.5$ is marked by the vertical red line. \textbf{b)}: The marginal $2$D posterior distributions of dynamic pressure against mass released for each fragmentation point (colors) identified from the USG light curve and mass remaining at peak dynamic pressure (black). Contours show the $1\sigma$, $2\sigma$ and $3\sigma$ bounds of the posterior distributions. Also marked on the plot are the dynamic pressure vs. mass released at the main fragmentation point (yellow dot) and at peak dynamic pressure (red dot) previously estimated by \citeA{brown_entry_2002}.
    }
    \label{fig:tagish_lake_lc}
\end{figure*}

In this work we model the USG light curve as seen in Figure \ref{fig:tagish_lake_lc}a), focusing on the two large fragmentations at $37$ and $32$ km altitude identified by \citeA{brown_entry_2002} and visible in the USG light curve. The posterior parameter estimates from our nested sampling-based model largely concur with previous results, with the main-burst dynamic pressure being less than a factor of two higher and peak dynamic pressure being a factor of three higher than the values estimated by \citeA{brown_entry_2002}.

% found that the object underwent extensive, continuous fragmentation from $50$ to $32$ km height, with the first major burst point at $37$ km and dynamic pressure of $0.7$ MPa, where $\sim26$ tons of the original $\sim56000$-ton mass remained.

\subsection{Mor\'{a}vka}\label{sec:moravka}

The daytime Mor\'{a}vka meteorite fall in the eastern Czech Republic on 6 May 2000 was accompanied by a bright fireball detected by USG sensors, seismic and infrasonic stations and by several serendipitous casual video recordings \cite{borovicka_moravka_2003-1}. The latter provided details of fragmentation, lateral fragment velocity spread and mass loss as a function of height. 
\citeA{neder_radionuclide_2001} and \citeA{borovicka_moravka_2003-2} determined the pre-impact mass of the object using several different techniques, producing a range from $230-11000$ kg with a best overall estimate of $1500 \pm 500$ kg \cite{borovicka_moravka_2003-2}.
Dynamical modeling of observed fragments by \citeA{borovicka_moravka_2003} using the video footage found that significant low-altitude gross fragmentation began at a height of $32.3$ km under dynamic pressure of $4.3$ MPa. At this point, the object had already been broken into four large fragments of masses $142$, $114$, $106$ and $77$ kg (determined based on their observed deceleration) which were released unobserved at higher altitude.

\begin{figure*}
\centering
\includegraphics[width=1.\linewidth]{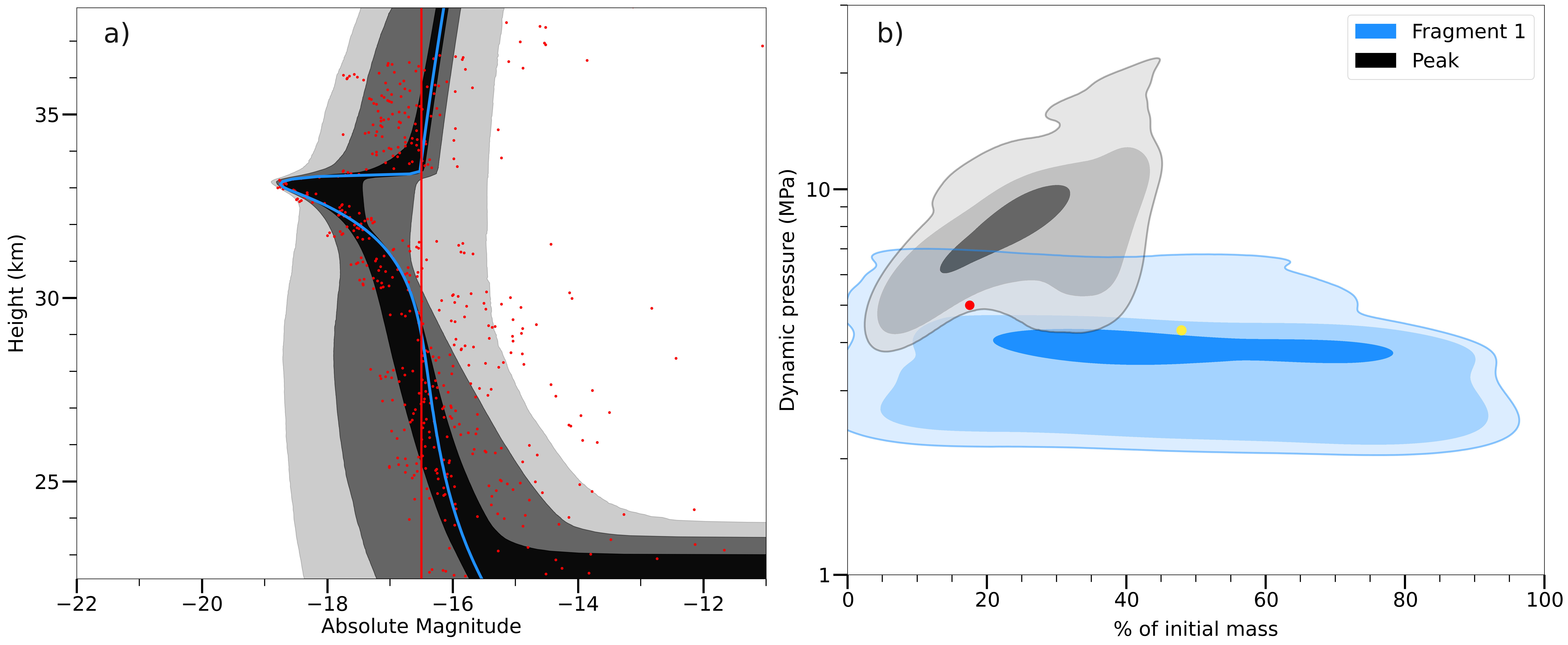}
    \caption{
    \textbf{a)}: The observed light curve and corresponding model fit for the 6 May 2000 Mor\'{a}vka fireball, similar to Figure \ref{fig:tagish_lake_lc}a). 
    % Here $1\sigma$, $2\sigma$ and $3\sigma$ uncertainties in the model fit are illustrated by the black shaded regions. The maximum log-likelihood solution obtained by nested sampling is plotted as the blue line. The intensity is given in units of absolute stellar magnitudes assuming a bolometric power of $3030$ W at zero magnitude. The detection limit of USG sensors at absolute magnitude $-16.5$ is marked by the vertical red line. 
    Evidence of extended luminosity possibly produced by dust left behind by the bolide after its final fragmentation is visible as an inflection in the light curve at $\sim31$ km.  \textbf{b)}: Comparison of the posterior distributions for dynamic pressure against mass released at the main fragmentation and mass remaining at peak dynamic pressure to previous estimates by \citeA{borovicka_moravka_2003}, similar to Figure \ref{fig:tagish_lake_lc}b). % Contours show the $1\sigma$, $2\sigma$ and $3\sigma$ bounds of the posterior distributions. 
    % Also marked on the plot are the pressure vs. mass released at the main fragmentation point (yellow) and at peak dynamic pressure (red) previously estimated by \citeA{borovicka_moravka_2003}.
    }
    \label{fig:moravka_lc}
\end{figure*}

The four initial fragments tracked by \citeA{borovicka_moravka_2003} then each fragmented again between $32.3$ and $29.3$ km, with the overall peak dynamic pressure of $5.0$ MPa reached by the $77$-kg fragment at $29.3$ km. Smaller fragmentation events were then subsequently tracked by \citeA{borovicka_moravka_2003} down to $24.0$ km.
The fraction of overall mass of the four initial fragments represented by the $77$-kg fragment is taken as our estimate for mass remaining at peak dynamic pressure.
% We take the mass released by the largest $142$-kg fragment observed as an estimate for the overall mass fraction released by the bolide at the main fragmentation point.

Here we instead model the USG light curve using only a single fragmentation event at $\sim33$ km where a large flare can be seen in Figure \ref{fig:moravka_lc}a), corresponding to the onset of fragmentation as identified by \citeA{borovicka_moravka_2003}. We do not model the other fragmentations as the rest of the USG light curve is close to the detection limit and consequently of low fidelity.
Comparing our nested sampling-based fit with the USG sensor data to the dynamical modeling results of \citeA{borovicka_moravka_2003}, we find that our results compare very favorably overall, with only our inferred peak dynamic pressure being higher than the $5.0$ MPa estimated by \citeA{borovicka_moravka_2003}. We suggest this may be due to residual luminosity detected by USG sensors from the bolide's sunlit stationary trail being mapped to light production at apparently lower altitudes. This can be seen in the Figure \ref{fig:moravka_lc}a) light curve but is not a real feature, as evidenced by the dynamical modeling of \citeA{borovicka_moravka_2003}.

\subsection{Park Forest}\label{sec:park_forest}

The Park Forest meteorite fell in the United States on 27 March 2003. It was accompanied by a bright fireball detected by USG sensors, seismic, infrasonic and acoustic microphones and by casual videos \cite{brown_orbit_2004}. 
Mass estimates with several different techniques by \citeA{brown_orbit_2004} yielded a pre-impact mass for the object ranging between $7000-21000$ kg with a nominal best estimate of $11000 \pm 3000$ kg.
Using the gross fragmentation model of \citeA{ceplecha_atmospheric_1993}, \citeA{brown_orbit_2004} also analyzed a fused light curve of USG sensor and ground-based video observations to determine fireball fragmentation behavior.
They identified three large fragmentation points at $37$, $29$ and $22$ km, with the main burst occurring at $29$ km under a dynamic pressure of $7$ MPa (possibly overestimated by $20-30\%$). \citeA{brown_orbit_2004} also determined using the gross fragmentation model that the overall peak dynamic pressure of $14$ MPa was reached at a height of $\sim21.3$ km, where $\sim 92.5$ kg of the $11000$ kg initial mass remained.
% with a bolide model including meteoroid porosity that $\sim150$ kg of the $7000$ kg initial mass reached the ground as meteorites, which we take as an estimate for the mass remaining at peak dynamic pressure. \noteic{elsewhere in the paper it estimates $\sim50$ kg out of $11000$ kg initial mass?}

\begin{figure*}
\centering
\includegraphics[width=1.\linewidth]{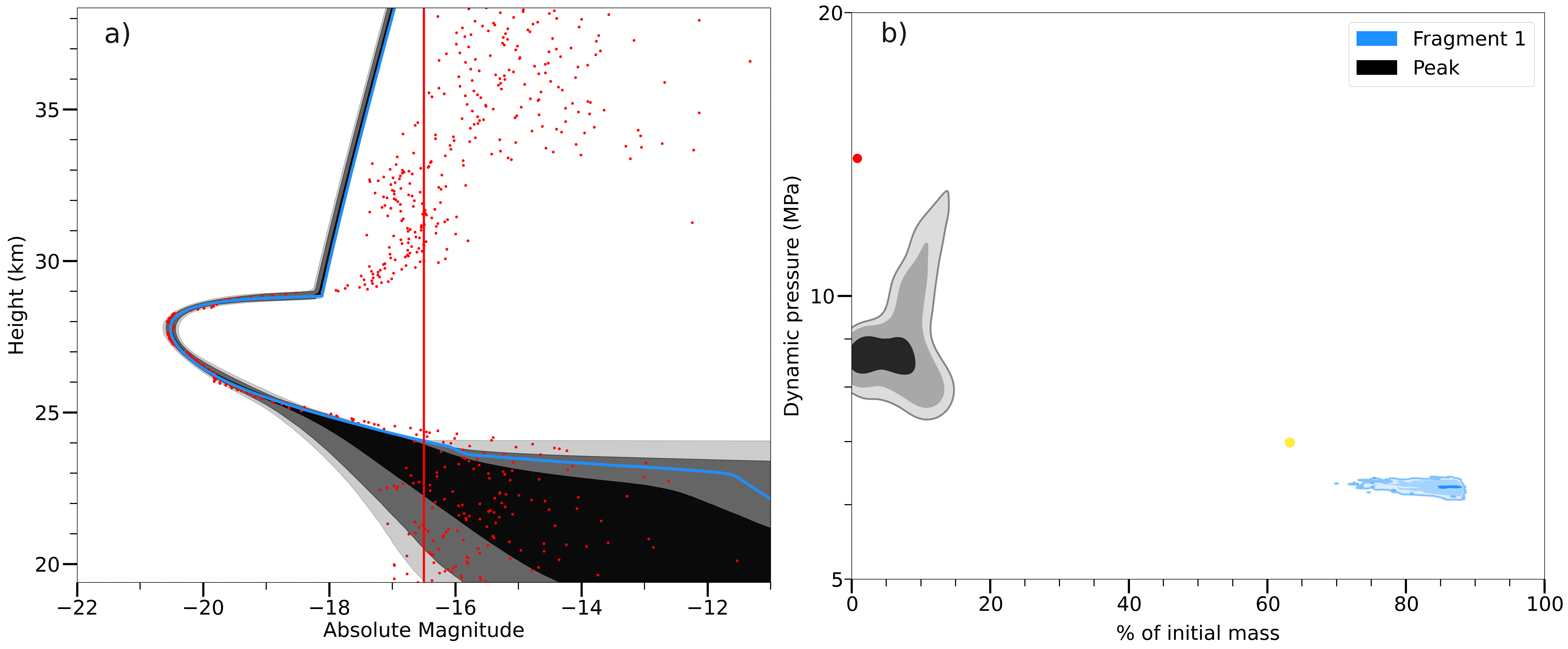}
    \caption{
    \textbf{a)}: The observed light curve and corresponding model fit for the 27 March 2003 Park Forest fireball, similar to Figure \ref{fig:tagish_lake_lc}a).
    % The fit light curve of intensity versus height plotted over the USG sensor-recorded fireball light curve (red) for the 27 March 2003 Park Forest meteorite-producing fireball. Here $1\sigma$, $2\sigma$ and $3\sigma$ uncertainties in the model fit are illustrated by the black shaded regions. The maximum log-likelihood solution obtained by nested sampling is plotted as the blue line. The intensity is given in units of absolute stellar magnitudes assuming a bolometric power of $3030$ W at zero magnitude. The detection limit of USG sensors at absolute magnitude $-16.5$ is marked by the vertical red line. 
    \textbf{b)}: 
    Comparison of the posterior distributions for dynamic pressure against mass released at the main fragmentation and mass remaining at peak dynamic pressure to previous estimates by \citeA{brown_orbit_2004}, similar to Figure \ref{fig:tagish_lake_lc}b).
    % The marginal $2$D posterior distributions of dynamic pressure against mass released for the main fragmentation point (blue) which was identified from the USG light curve and mass remaining at peak dynamic pressure (black). Contours show the $1\sigma$, $2\sigma$ and $3\sigma$ bounds of the posterior distributions. Also marked on the plot are the pressure vs. mass released at the main fragmentation point (yellow) and at peak dynamic pressure (red) previously estimated by \citeA{brown_orbit_2004}.
    }
    \label{fig:park_forest_lc}
\end{figure*}

Here we model the USG light curve considering only the main burst at $29$ km identified by \citeA{brown_orbit_2004} and shown in Figure \ref{fig:park_forest_lc}a), as the other two fragmentations occurred close to the USG sensor detection limit. Our nested sampling-based results for initial mass, dynamic pressure at main burst, and mass remaining at peak dynamic pressure are consistent with the estimates reported by \citeA{brown_orbit_2004}. The mass released at the main fragmentation is somewhat larger than previous nominal estimates by \citeA{brown_orbit_2004}, while the main-burst and peak dynamic pressures are somewhat smaller. The former discrepancy is likely because we do not consider the first large fragmentation at $37$ km identified by \citeA{brown_orbit_2004} in this work (where some mass would have been released), while the latter could be due to differences between the simpler gross fragmentation model used by \citeA{brown_orbit_2004} in their analysis compared to the semi-empirical model of \citeA{borovicka_kosice_2013} employed in this work.

% Our nested sampling-based results for initial mass, dynamic pressure at the main burst, and mass remaining at peak dynamic pressure are consistent with the estimates reported by \citeA{brown_orbit_2004}.

\subsection{Ko\v{s}ice}\label{sec:kosice}

The Ko\v{s}ice meteorite fall in eastern Slovakia on 28 February 2010 was accompanied by a bright fireball detected by USG sensors, radiometers of the ground-based European Fireball Network (EN), seismic and infrasonic stations, and surveillance cameras in Hungary \cite{borovicka_kosice_2013}. 
Using the radiometric light curve, \citeA{borovicka_kosice_2013} determined a pre-impact mass of $3500$ kg citing  a factor of $3$ uncertainty in this value. The main fragmentation occurred at a height of $38.9$ km under dynamic pressure of $0.97$ MPa. \citeA{borovicka_kosice_2013} also identified a second large fragmentation of the main body at $29.3$ km. The peak dynamic pressure of $5.9$ MPa was reached by an $18.1$-kg fragment at $21.6$ km.

\begin{figure*}
\centering
\includegraphics[width=1.\linewidth]{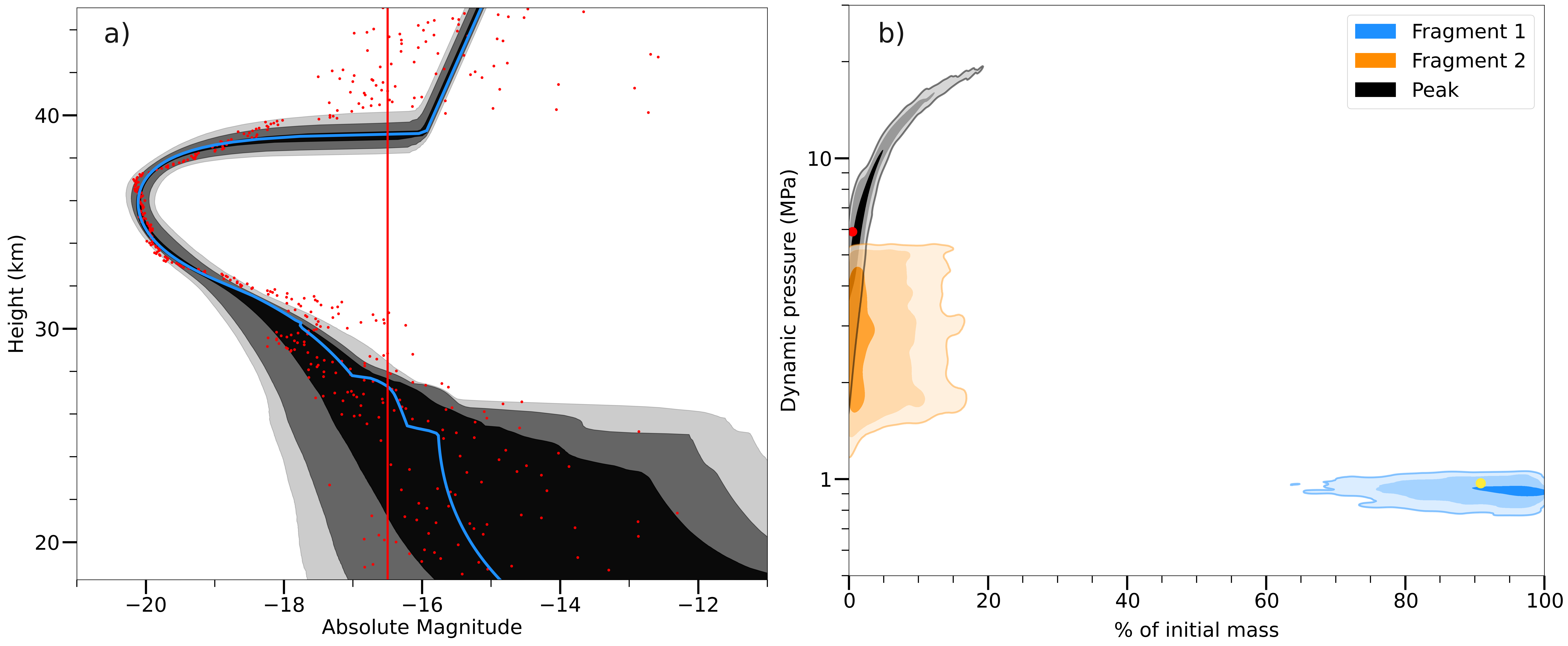}
    \caption{
    \textbf{a)}: The observed light curve and corresponding model fit for the 28 February 2010 Ko\v{s}ice fireball, similar to Figure \ref{fig:tagish_lake_lc}a).
    % The fit light curve of intensity versus height plotted over the USG sensor-recorded fireball light curve (red) for the 28 February 2010 Ko\v{s}ice meteorite-producing fireball. Here $1\sigma$, $2\sigma$ and $3\sigma$ uncertainties in the model fit are illustrated by the black shaded regions. The maximum log-likelihood solution obtained by nested sampling is plotted as the blue line. The intensity is given in units of absolute stellar magnitudes assuming a bolometric power of $3030$ W at zero magnitude. The detection limit of USG sensors at absolute magnitude $-16.5$ is marked by the vertical red line. 
    \textbf{b)}: Comparison of the posterior distributions for dynamic pressure against mass released at the main fragmentation and mass remaining at peak dynamic pressure to previous estimates by \citeA{borovicka_kosice_2013}, similar to Figure \ref{fig:tagish_lake_lc}b).
    % The marginal $2$D posterior distributions of dynamic pressure against mass released for each fragmentation point (colors) which were identified from the USG light cruve and mass remaining at peak dynamic pressure (black). Contours show the $1\sigma$, $2\sigma$ and $3\sigma$ bounds of the posterior distributions. Also marked on the plot are the pressure vs. mass released at the main fragmentation point (yellow) and at peak dynamic pressure (red) previously estimated by \citeA{borovicka_kosice_2013}.
    }
    \label{fig:kosice_lc}
\end{figure*}

In our model for this work only the two fragmentation points at $38.9$ and $29.3$ km are considered as they are clearly visible in the USG sensor light curve shown in Figure \ref{fig:kosice_lc}a). We combine the three smaller fragmentations of previously released fragments at $36.6$, $34.8$ and $33.7$ km identified by \citeA{borovicka_kosice_2013} into the main fragmentation at $38.9$ km here, totaling $3175$ kg released from the initial mass of $3496$ kg. % 2750 is adding up EFs, 3175 is adding up all parent masses
Our posterior parameter estimates are consistent with the results obtained by \citeA{borovicka_kosice_2013} with the exception of the initial mass, which we find to be substantially larger in part because the corrected USG lightcurve is almost two magnitudes brighter than the inferred peak brightness from ground-based detectors. However, as noted by \citeA{borovicka_kosice_2013}, the ground-based brightness is uncertain to roughly one stellar magnitude so this difference is not unexpected. 

\subsection{Romanian Fireball}\label{sec:romania}

A superbolide occurred over central Romania on 7 January 2015, from which no meteorites were recovered \cite{borovicka_january_2017}. The fireball was observed by USG sensors, EN all-sky cameras and radiometers located in the Czech Republic and Slovakia, and surveillance cameras in Romania. Analysis by \citeA{borovicka_january_2017} determined a pre-impact mass of $\sim4500$ kg with uncertainty of $\pm50\%$. Modeling the radiometric light curve, \citeA{borovicka_january_2017} identified an initial flare releasing $30$ kg of mass at $53.5$ km, followed by severe fragmentation beginning at $48$ km height and dynamic pressure of $0.9$ MPa and occurring almost continuously down to $42$ km, releasing a total of $4050$ kg. The remaining body started eroding at $40.9$ km and completely disintegrated into dust around $38$ km after reaching a peak dynamic pressure of $3.0$ MPa.

\begin{figure*}
\centering
\includegraphics[width=1.\linewidth]{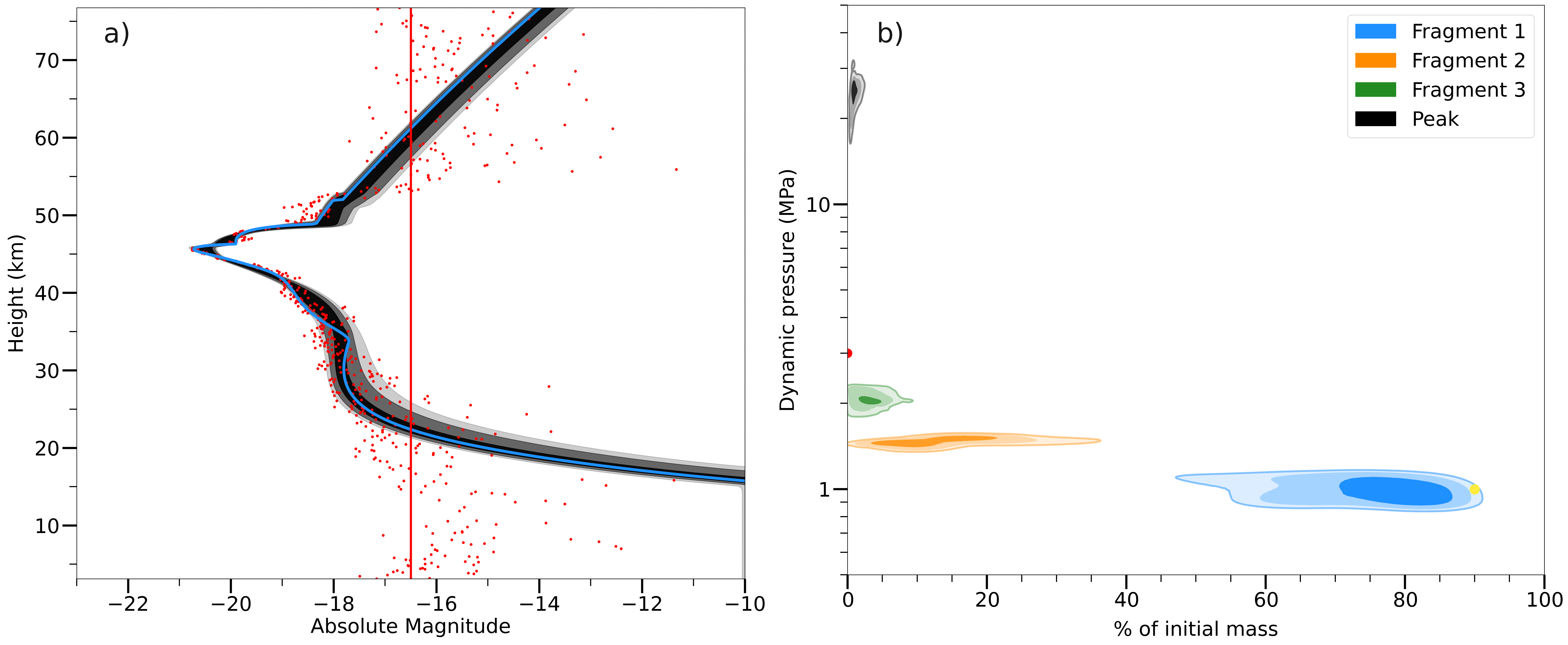}
    \caption{
    \textbf{a)}: The observed light curve and corresponding model fit for the 7 January 2015 Romanian superbolide, similar to Figure \ref{fig:tagish_lake_lc}a).
    % The fit light curve of intensity versus height plotted over the USG sensor-recorded light curve (red) for the 7 January 2015 Romanian superbolide. Here $1\sigma$, $2\sigma$ and $3\sigma$ uncertainties in the model fit are illustrated by the black shaded regions. The maximum log-likelihood solution obtained by nested sampling is plotted as the blue line. The intensity is given in units of absolute stellar magnitudes assuming a bolometric power of $3030$ W at zero magnitude. The detection limit of USG sensors at absolute magnitude $-16.5$ is marked by the vertical red line.
    Similar to Figure \ref{fig:moravka_lc}a), extended luminosity appears as an inflection in the light curve at $\sim35$ km.
    % Evidence of extended luminosity possibly produced by dust left behind by the bolide after its final fragmentation is visible as an inflection in the light curve at $\sim35$ km.    
    \textbf{b)}: Comparison of the posterior distributions for dynamic pressure against mass released at main fragmentation and mass remaining at peak dynamic pressure to previous estimates by \citeA{borovicka_january_2017}, similar to Figure \ref{fig:tagish_lake_lc}b).
    % The marginal $2$D posterior distributions of dynamic pressure against mass released for each fragmentation point (colors) which were identified from the USG light curve and mass remaining at peak dynamic pressure (black). Contours show the $1\sigma$, $2\sigma$ and $3\sigma$ bounds of the posterior distributions. Also marked on the plot are the pressure vs. mass released at the main fragmentation point (yellow) and at peak dynamic pressure (red) previously estimated by \citeA{borovicka_january_2017}.
    }
    \label{fig:romania_lc}
\end{figure*}

Here we model the bolide considering the initial flare at $53.5$ km and the main fragmentation at $48$ km identified by \citeA{borovicka_january_2017} and apparent in the USG light curve. The subsequent individual fragmentations are all combined into a single fragmentation at $\sim 46$ km to simplify the model. The USG light curve and our resulting fit can be seen in Figure \ref{fig:romania_lc}a). Our estimate for the dynamic pressure and mass released at the main fragmentation point match well with the values estimated by \citeA{borovicka_january_2017}, while our peak dynamic pressure is an order of magnitude higher. Similar to Mor\'{a}vka, this discrepancy is likely due to ongoing trail emission being detected by USG sensors mapping to apparent (but non-existent) low-altitude luminosity as seen in Figure \ref{fig:romania_lc}a). We model this extended light for consistency here (treating the USG light curve as the only information we have for this event) but \citeA{borovicka_january_2017} do not, as they recognized this ``shelf" of persistent luminosity as being due to the fixed bolide trail which they were able to directly resolve.

\subsection{Sari\c{c}i\c{c}ek}\label{sec:saricicek}

The Sari\c{c}i\c{c}ek meteorite fall in Turkey on 2 September 2015 was accompanied by a bright fireball detected by USG sensors, seismic sensors, and security cameras in Turkey \cite{unsalan_saricicek_2019}. Analysis by \citeA{unsalan_saricicek_2019} determined the object's pre-impact diameter to be $1.0\pm0.2$ m, which when combined with the $2910\pm20$ kg m$^{-3}$ bulk density of recovered meteorites yields a pre-impact mass estimate of $775-2651$ kg. Using the triggered progressive fragmentation model of \citeA{revelle_recent_2004} to model the security camera light curve, \citeA{unsalan_saricicek_2019} identified flares at heights $36.4$, $33.0$, $31.0$, and $27.4$ km, with dynamic pressures of $1.00$ MPa at the first (and largest) flare and $3.99$ MPa at the last flare, where the object completely disrupted. We take this last value to be the peak dynamic pressure and the total mass of $24.78$ kg recovered meteorites \cite{unsalan_saricicek_2019} to be an estimate for the mass remaining at peak dynamic pressure. The masses released at each fragmentation are not reported.

\begin{figure*}
\centering
\includegraphics[width=1.\linewidth]{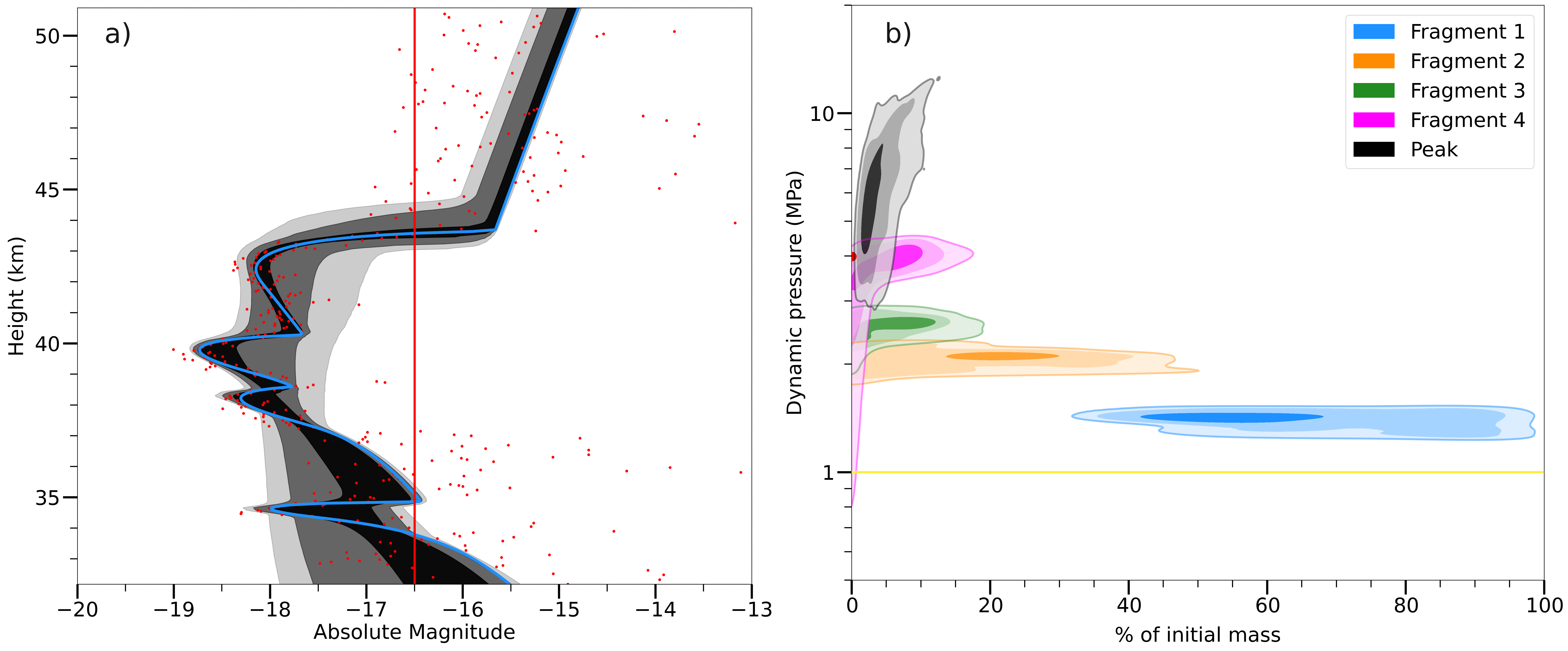}
    \caption{
    \textbf{a)}: The observed light curve and corresponding model fit for the 2 September 2015 Sari\c{c}i\c{c}ek fireball, similar to Figure \ref{fig:tagish_lake_lc}a).    
    % The fit light curve of intensity versus height plotted over the USG sensor-recorded fireball light curve (red) for the 2 September 2015 Sari\c{c}i\c{c}ek meteorite-producing fireball. Here $1\sigma$, $2\sigma$ and $3\sigma$ uncertainties in the model fit are illustrated by the black shaded regions. The maximum log-likelihood solution obtained by nested sampling is plotted as the blue line. The intensity is given in units of absolute stellar magnitudes assuming a bolometric power of $3030$ W at zero magnitude. The detection limit of USG sensors at absolute magnitude $-16.5$ is marked by the vertical red line. 
    \textbf{b)}: Comparison of the posterior distributions for dynamic pressure against mass released at main fragmentation and mass remaining at peak dynamic pressure to previous estimates by \citeA{unsalan_saricicek_2019}, similar to Figure \ref{fig:tagish_lake_lc}b). Note that the estimated dynamic pressure at the main fragmentation point by \citeA{unsalan_saricicek_2019} is indicated by a line rather than a dot in this case as the released mass is not reported.
    % The marginal $2$D posterior distributions of dynamic pressure against mass released for each fragmentation point (colors) which were identified from the USG light curve and mass remaining at peak dynamic pressure (black). Contours show the $1\sigma$, $2\sigma$ and $3\sigma$ bounds of the posterior distributions. The yellow line shows the estimated dynamic pressure at the main fragmentation point by \citeA{unsalan_saricicek_2019} (the released mass is not reported). The red dot shows the pressure vs. mass remaining at peak dynamic pressure estimated by \citeA{unsalan_saricicek_2019}.
    }
    \label{fig:saricicek_lc}
\end{figure*}

In this work we model the Sari\c{c}i\c{c}ek bolide light curve as having four fragmentation events as evidenced by the USG light curve. These are common with the four flares identified by \citeA{unsalan_saricicek_2019}. The USG light curve and our resulting fit obtained using nested sampling is shown in Figure \ref{fig:saricicek_lc}a). We note that our dynamic pressure at the main fragmentation is somewhat higher than that of \citeA{unsalan_saricicek_2019}, while our other estimates are generally consistent. This discrepancy is likely due to the difference in the height of peak brightness reported by USG sensors ($39.8$ km) compared with that determined via the security camera footage by \citeA{unsalan_saricicek_2019} ($36.2$ km); the four flares are instead at heights $\sim44.9$ km, $\sim40.5$ km, $\sim38.6$ km, and $\sim35.0$ km in our model.

\subsection{Flensburg}\label{sec:flensburg}

The Flensburg C1-ungrouped meteorite fall occurred in northern Germany on 12 September 2019 \cite{bischoff_old_2021, borovicka_trajectory_2021}. The associated daytime bolide was observed by USG sensors, the ground-based AllSky6 meteor observing camera system \cite{hankey_all-sky-6_2020} and dashboard cameras. \citeA{borovicka_trajectory_2021} obtained a pre-impact mass range of $10,000-22,000$ kg for the object. The lower mass estimate was derived using the USG-reported impact energy, while the upper estimate was derived by fitting an approximate ground-based light curve normalized to the total USG sensor energy with the semi-empirical model of \citeA{borovicka_kosice_2013} using a similar method to that of \citeA{borovicka_maribo_2019} for the CM2 carbonaceous chondrite Maribo. Based on detailed modeling of the AllSky6 and dashcam light curves, \citeA{borovicka_trajectory_2021} also identified the main fragmentation of the object at a height of $45.5$ km and dynamic pressure of $0.7$ MPa, releasing $15000$ kg from the $22000$ kg main body. % (J. Borovi\v{c}ka, personal communcation, 2025). 
This was followed by smaller fragmentations at $42.5$ and $37$ km, the latter of which saw the object reach its peak dynamic pressure of $2.0$ MPa where an estimated $\sim3\%$ of initial mass remained.

\begin{figure*}
\centering
\includegraphics[width=1.\linewidth]{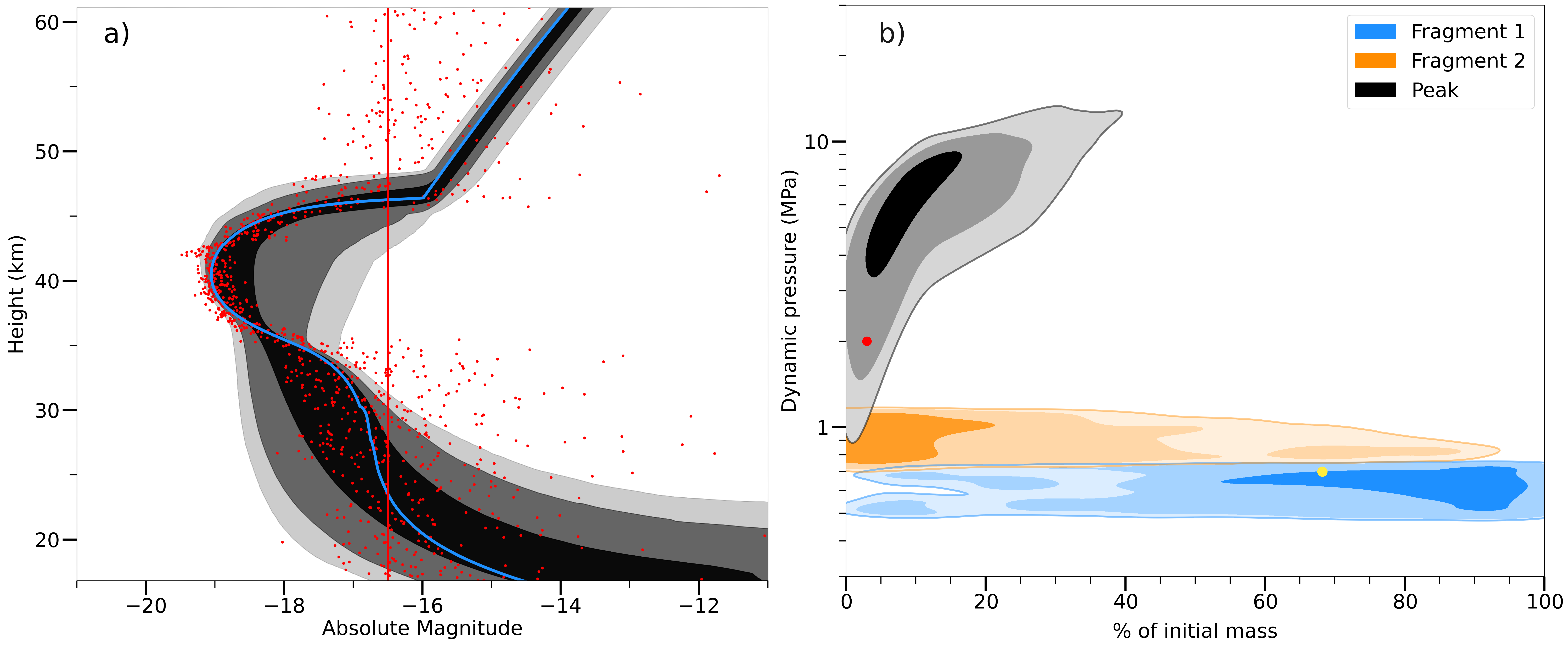}
    \caption{
    \textbf{a)}: The observed light curve and corresponding model fit for the 12 September 2019 Flensburg fireball, similar to Figure \ref{fig:tagish_lake_lc}a). Similar to Figure \ref{fig:moravka_lc}a), extended luminosity appears as an inflection in the light curve at $\sim35$ km.
    % The fit light curve of intensity versus height plotted over the USG sensor-recorded fireball light curve (red) for the daytime 12 September 2019 Flensburg meteorite-producing fireball. Here $1\sigma$, $2\sigma$ and $3\sigma$ uncertainties in the model fit are illustrated by the black shaded regions. The maximum log-likelihood solution obtained by nested sampling is plotted as the blue line. The intensity is given in units of absolute stellar magnitudes assuming a bolometric power of $3030$ W at zero magnitude. The detection limit of USG sensors at absolute magnitude $-16.5$ is marked by the vertical red line. 
    % Evidence of extended luminosity possibly produced by dust left behind by the bolide after its final fragmentation is visible as an inflection in the light curve at $\sim35$ km. 
    \textbf{b)}: Comparison of the posterior distributions for dynamic pressure against mass released at main fragmentation and mass remaining at peak dynamic pressure to previous estimates by \citeA{borovicka_trajectory_2021}, similar to Figure \ref{fig:tagish_lake_lc}b).
    % The marginal $2$D posterior distributions of dynamic pressure against mass released for each fragmentation point (colors) and mass remaining at peak dynamic pressure (black). Contours show the $1\sigma$, $2\sigma$ and $3\sigma$ bounds of the posterior distributions. Also marked on the plot are the pressure vs. mass released at the main fragmentation point (yellow) and at peak dynamic pressure (red) previously estimated by \citeA{borovicka_trajectory_2021}.
    }
    \label{fig:flensburg_lc}
\end{figure*}

We model the light curve here considering only the two fragmentations found by \citeA{borovicka_trajectory_2021} at $45.5$ and $42.5$ km, as the final fragmentation at $37.5$ km is not visible in the USG light curve (shown in Figure \ref{fig:flensburg_lc}a)).
Our inferred estimates of initial mass, dynamic pressure and mass released at the main fragmentation point from the USG sensor light curve are broadly consistent with previous work by \citeA{borovicka_trajectory_2021}, while our inferred peak dynamic pressure and mass remaining at peak pressure are again substantially higher. As with the Mor\'{a}vka and the Romanian bolides, this may be due to lingering emission in the daytime sky producing an apparent low-altitude luminous trail (as seen in Figure \ref{fig:flensburg_lc}a)) detected by the USG sensors. 

\subsection{Discussion}

Overall, we find that for these seven validation bolide cases our initial mass estimates are similar to those estimated in the literature, noting that differences in the USG versus ground-based light curves and assumptions in luminous efficiency can easily produce deviations of a factor of several in the initial mass. For those fireballs for which detailed modelling was available, we find that our dynamic pressure where the majority of the mass is released is most similar to our results. We therefore suggest that our method is most robust at estimating the dynamic pressure and associated mass loss at primary fragmentation. 

We note that for the Mor\'{a}vka, Romania and Flensburg bolides, our inferred peak dynamic pressure is substantially higher than previous estimates. This is likely due to the common observations by USG sensors of an extended, stationary luminous trail produced by optically thick hot vapor and/or residual dust \cite{borovicka_january_2017}. In all three of these cases, this feature appears as an inflection in the light curve after the main or final fragmentation. For daytime fireballs such as Mor\'{a}vka and Flensburg, reflected sunlight from the debris cloud may also be a contributor as this effect has been previously identified as causing long ``tails" in USG sensor light curves \cite{tagliaferri_analysis_1995}.

A major result of these comparisons is that caution should be exercised in fitting or interpreting USG sensor light curves past their point of peak brightness. It appears in many cases that extended emission is a common feature and may result in inflated peak dynamic pressures or total optical energies.

\section{Decameter Impactors}\label{sec:decameter_impactors}

Here we apply our nested sampling method to $13$ decameter-size impactors, consisting of all impactors identified in Table $1$ of \citeA{chow_decameter-sized_2025} with the exception of the 7 February 2022 South Atlantic fireball, which is excluded from our analysis. We excluded this event as the USG light curve shows an unphysical rise in post-peak brightness consistent with a signature of contamination from a persistent trail. This suggests its energy is overestimated. Indeed, infrasound records from six stations which show a clear signal from the bolide produce an average period of $6.8$ sec. The multi-station average period-yield relationship of \citeA{ens_infrasound_2012} therefore suggests a source yield estimate below $3$ kilotons of TNT ($1$ kT TNT $=4.184\times10^{12}$ J), further supporting our choice to remove this event from consideration as it is likely well below the $7.5$m diameter threshold of our survey.

\begin{figure*}
\centering
\includegraphics[width=1.\linewidth]{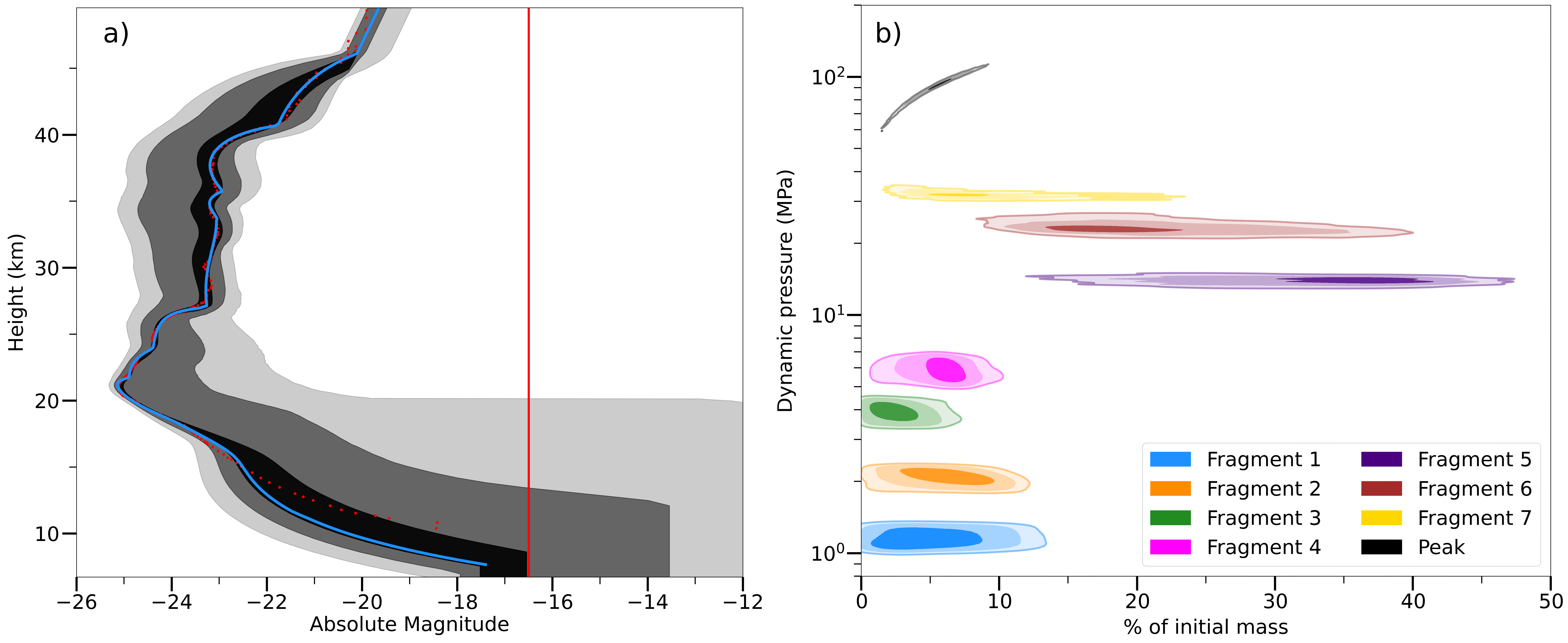}
    \caption{
    \textbf{a)}: The observed light curve and corresponding model fit for the 1 February 1994 Marshall Islands fireball, similar to Figure \ref{fig:tagish_lake_lc}a).   
    % The fit light curve of intensity versus height plotted over the USG sensor-recorded fireball light curve (red dots) for the 1 February 1994 Marshall Islands fireball. Here $1\sigma$, $2\sigma$ and $3\sigma$ uncertainties are illustrated by the black shaded regions. The maximum log-likelihood solution obtained by nested sampling is plotted as the blue line. The intensity is given in units of absolute stellar magnitudes assuming
    % a bolometric power of $3030$ W at zero magnitude. The detection limit of USG sensors at absolute magnitude $-16.5$ is marked by the vertical red line. 
    \textbf{b)}: The posterior distributions for dynamic pressure against mass released for each fragmentation point and mass remaining at peak dynamic pressure, similar to Figure \ref{fig:tagish_lake_lc}b).
    % The marginal $2$D posterior distributions of dynamic pressure against mass released for each fragmentation point (colors) and mass remaining at peak dynamic pressure (black). Contours show the $1\sigma$, $2\sigma$ and $3\sigma$ bounds of the posterior distributions.
    }
    \label{fig:marshall_islands_lc}
\end{figure*}

\begin{figure*}
    \centering
    \includegraphics[width=1.\linewidth]{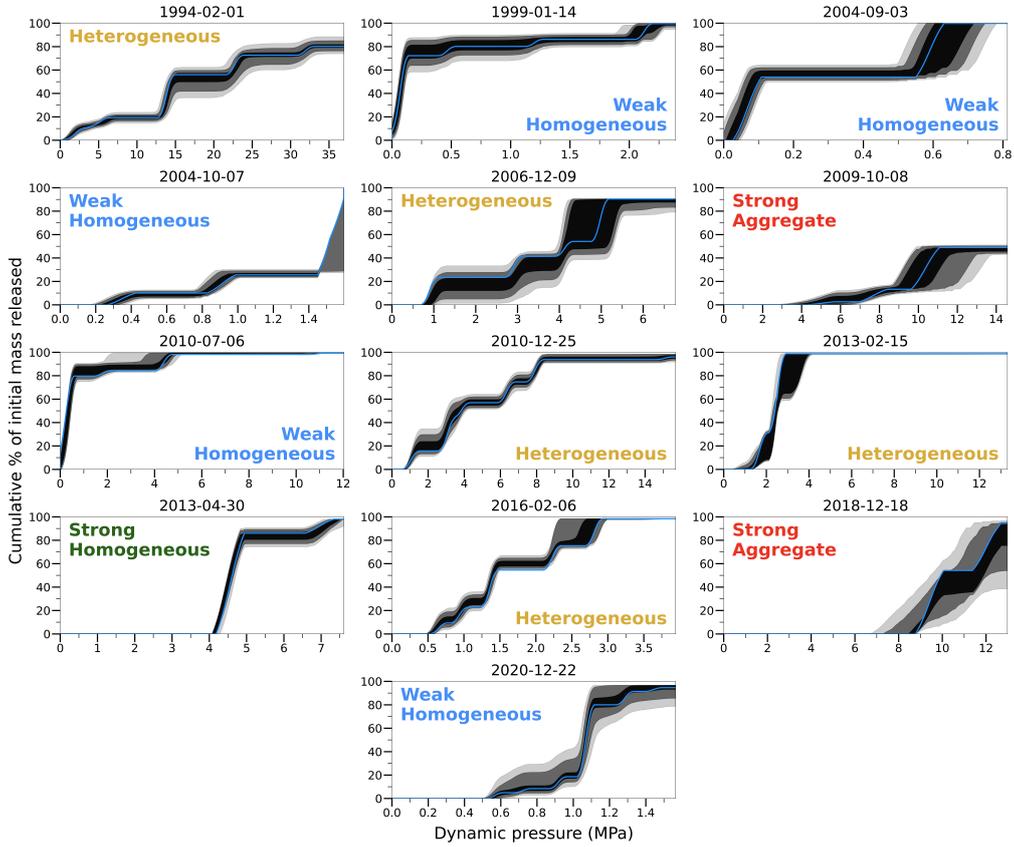}
    \caption{A staircase plot showing the cumulative mass released (expressed as a percentage of the initial mass) by $13$ decameter-size impactors as a function of dynamic pressure, in MPa. Each object is identified by its UTC date (YYYY-MM-DD) of impact. The structural class of each object is also indicated. The black shaded regions illustrate the $1\sigma$, $2\sigma$ and $3\sigma$ distributions of the cumulative mass released at a given dynamic pressure. The cumulative mass release corresponding to the maximum log-likelihood solution for each impactor is plotted in blue.}
    \label{fig:mass_cdfs}
\end{figure*}

As for the validation events, the maximum log-likelihood solution as well as the $5\%$, $50\%$ and $95\%$ quantiles of the $1$D posterior distributions of relevant physical parameters for each event are recorded. These results are summarized in Table \ref{tab:decameter}. Figure \ref{fig:marshall_islands_lc}a) shows an example light curve for one of the decameter impactors (the 1 February 1994 Marshall Islands fireball) fit using our procedure, while Figure \ref{fig:marshall_islands_lc}b) shows the marginal $2$D posterior distributions of dynamic pressure vs. mass released for each fragment and at peak dynamic pressure obtained using nested sampling. The fit light curves and marginal $2$D posterior distributions of dynamic pressure vs. mass released for the other $12$ decameter impactors are shown in Figures S$1$-S$24$ in Supporting Information S$1$. The cumulative mass released for each object as a function of dynamic pressure is shown in Figure \ref{fig:mass_cdfs}.

\begin{table*}
\centering
\makebox[1 \textwidth][r]{       %centering table
\resizebox{1.2 \textwidth}{!}{   %resize table
\begin{tabular}{|l|ccc|ccc|ccc|ccc|ccc|l|}
\hline
\textbf{UTC Date}
% & \multicolumn{3}{c|}{{$\bm{m_\mathrm{init}}$ (kg)}}
& \multicolumn{3}{c|}{{$\bm{m_\mathrm{init}}$ ($10^5$kg)}}
& \multicolumn{3}{c|}{{$\bm{P_\mathrm{main}}$ (MPa)}}
& \multicolumn{3}{c|}{{$\bm{m_\mathrm{main}}$ (\% of $m_\mathrm{init}$)}}
& \multicolumn{3}{c|}{{$\bm{P_\mathrm{peak}}$ (MPa)}}
& \multicolumn{3}{c|}{{$\bm{m_\mathrm{peak}}$ (\% of $m_\mathrm{init}$)}} 
& \textbf{Structural Class} \\
(YYYY-MM-DD) &
$5\%$ & $50\%$ & $95\%$ &
$5\%$ & $50\%$ & $95\%$ &
$5\%$ & $50\%$ & $95\%$ &
$5\%$ & $50\%$ & $95\%$ &
$5\%$ & $50\%$ & $95\%$ &
\\
\hline
% 1994-02-01 max. $\log\mathcal{L}$ & \multicolumn{3}{c|}{$471244$} & \multicolumn{3}{c|}{$13.81$} & \multicolumn{3}{c|}{$36.77$} & \multicolumn{3}{c|}{$93.99$} & \multicolumn{3}{c|}{$5.82$} \\
1994-02-01 max. $\log\mathcal{L}$ & \multicolumn{3}{c|}{$4.71$} & \multicolumn{3}{c|}{$13.8$} & \multicolumn{3}{c|}{$36.8$} & \multicolumn{3}{c|}{$94.0$} & \multicolumn{3}{c|}{$5.82$} & Heterogeneous \\
% 1994-02-01 quantiles  & $454297$ & $474209$ & $492726$ & $13.45$ & $14.16$ & $14.54$ & $22.10$ & $35.76$ & $41.08$ & $79.64$ & $93.21$ & $103.38$ & $3.61$ & $5.69$ & $7.47$ \\
1994-02-01 quantiles  & $4.54$ & $4.74$ & $4.92$ & $13.5$ & $14.2$ & $14.5$ & $22.1$ & $35.8$ & $41.1$ & $79.6$ & $93.2$ & $103$ & $3.61$ & $5.69$ & $7.47$ & \\
% cut out 10k/~35k
\hline
% 1999-01-14 max. $\log\mathcal{L}$ & \multicolumn{3}{c|}{$408246$} & \multicolumn{3}{c|}{$0.067$} & \multicolumn{3}{c|}{$65.7$} & \multicolumn{3}{c|}{$5.48$} & \multicolumn{3}{c|}{$0.031$} \\
% 1999-01-14 quantiles & $390499$ & $406441$ & $424616$ & $0.050$ & $0.067$ & $0.094$ & 
% $56.82$ & $70.32$ & $81.47$ & $5.20$ & $11.81$ & $25.10$ & $0.021$ & $0.28$ & $1.89$ \\
1999-01-14 max. $\log\mathcal{L}$ & \multicolumn{3}{c|}{$4.08$} & \multicolumn{3}{c|}{$0.0666$} & \multicolumn{3}{c|}{$65.7$} & \multicolumn{3}{c|}{$5.48$} & \multicolumn{3}{c|}{$0.0307$} & Weak Homogeneous \\
1999-01-14 quantiles & $3.90$ & $4.06$ & $4.25$ & $0.0501$ & $0.0666$ & $0.0939$ & 
$56.8$ & $70.3$ & $81.5$ & $5.20$ & $11.8$ & $25.1$ & $0.0211$ & $0.282$ & $1.89$ & \\  % check mpeak
% cut out 11k/~23k
\hline
% 2004-09-03 max. $\log\mathcal{L}$ & \multicolumn{3}{c|}{$715326$} & \multicolumn{3}{c|}{$0.068$} & \multicolumn{3}{c|}{$54.0$} & \multicolumn{3}{c|}{$2.14$} & \multicolumn{3}{c|}{$0.0056$} \\
% 2004-09-03 quantiles & $697317$ & $717919$ & $741180$ & $0.035$ & $0.059$ & $0.081$ & 
% $51.24$ & $56.42$ & $62.13$ & $1.88$ & $2.07$ & $2.65$ & $0.0038$ & $0.0063$ & $0.012$ \\
2004-09-03 max. $\log\mathcal{L}$ & \multicolumn{3}{c|}{$7.15$} & \multicolumn{3}{c|}{$0.0677$} & \multicolumn{3}{c|}{$54.0$} & \multicolumn{3}{c|}{$2.14$} & \multicolumn{3}{c|}{$0.00561$} & Weak Homogeneous \\
2004-09-03 quantiles & $6.97$ & $7.18$ & $7.41$ & $0.0355$ & $0.0590$ & $0.0812$ & $51.2$ & $56.4$ & $62.1$ & $1.88$ & $2.07$ & $2.65$ & $0.00382$ & $0.00637$ & $0.0122$ & \\
% cut out 12k/~25-26k
\hline
% 2004-10-07 max. $\log\mathcal{L}$ & \multicolumn{3}{c|}{$455306$} & \multicolumn{3}{c|}{$1.53$} & \multicolumn{3}{c|}{$73.49$} & \multicolumn{3}{c|}{$4.64$} & \multicolumn{3}{c|}{$0.054$} \\
% 2004-10-07 quantiles & $447415$ & $454544$ & $461978$ & $1.53$ & $1.53$ & $1.60$ & $69.93$ & $73.21$ & $75.31$ & $4.62$ & $4.69$ & $4.96$ & $0.034$ & $0.063$ & $0.10$ \\
2004-10-07 max. $\log\mathcal{L}$ & \multicolumn{3}{c|}{$4.55$} & \multicolumn{3}{c|}{$1.53$} & \multicolumn{3}{c|}{$73.5$} & \multicolumn{3}{c|}{$4.64$} & \multicolumn{3}{c|}{$0.054$} & Weak Homogeneous \\
2004-10-07 quantiles & $4.47$ & $4.55$ & $4.62$ & $1.53$ & $1.53$ & $1.60$ & $70.0$ & $73.2$ & $75.3$ & $4.62$ & $4.69$ & $4.96$ & $0.0344$ & $0.0626$ & $0.101$ & \\
% cut out 15k/~29k
\hline
2006-12-09 max. $\log\mathcal{L}$ & \multicolumn{3}{c|}{$3.65$} & \multicolumn{3}{c|}{$5.26$} & \multicolumn{3}{c|}{$35.5$} & \multicolumn{3}{c|}{$5.64$} & \multicolumn{3}{c|}{$9.41$} & Heterogeneous \\
2006-12-09 quantiles & $3.25$ & $3.58$ & $3.99$ & $5.24$ & $5.28$ & $5.36$ & $29.7$ & $37.2$ & $44.1$ & $5.51$ & $5.64$ & $6.25$ & $2.60$ & $4.63$ & $9.11$ & \\
%cut out 20k out of ~52600
\hline
% 2009-10-08 max. $\log\mathcal{L}$* & \multicolumn{3}{c|}{$622598$} & \multicolumn{3}{c|}{$10.36$} & \multicolumn{3}{c|}{$35.68$} & \multicolumn{3}{c|}{$185.82$} & \multicolumn{3}{c|}{$26.14$} \\
% 2009-10-08 quantiles & $614900$ & $624357$ & $635466$ & $9.18$ & $10.05$ & $12.46$ & $30.33$ & $34.37$ & $37.69$ & $183.25$ & $188.27$ & $193.63$ & $24.74$ & $25.85$ & $27.22$ \\
2009-10-08 max. $\log\mathcal{L}$* & \multicolumn{3}{c|}{$6.23$} & \multicolumn{3}{c|}{$10.4$} & \multicolumn{3}{c|}{$35.7$} & \multicolumn{3}{c|}{$186$} & \multicolumn{3}{c|}{$26.1$} & Strong Aggregate \\
2009-10-08 quantiles* & $6.15$ & $6.24$ & $6.35$ & $9.18$ & $10.0$ & $12.5$ & $30.3$ & $34.4$ & $37.7$ & $183$ & $188$ & $194$ & $24.7$ & $25.8$ & $27.2$ & \\
% cut out 10k/~23k
\hline
% 2010-07-06 max. $\log\mathcal{L}$ & \multicolumn{3}{c|}{$576573$} & \multicolumn{3}{c|}{$0.22$} & \multicolumn{3}{c|}{$79.68$} & \multicolumn{3}{c|}{$11.91$} & \multicolumn{3}{c|}{$0.98$} \\
% 2010-07-06 quantiles & $576222$ & $605123$ & $625617$ & $0.23$ & $0.30$ & $0.40$ & $76.97$ & $85.20$ & $89.67$ & $7.54$ & $8.27$ & $11.21$ & $0.35$ & $0.67$ & $1.23$ \\
2010-07-06 max. $\log\mathcal{L}$ & \multicolumn{3}{c|}{$5.77$} & \multicolumn{3}{c|}{$0.218$} & \multicolumn{3}{c|}{$79.7$} & \multicolumn{3}{c|}{$11.9$} & \multicolumn{3}{c|}{$0.984$} & Weak Homogeneous \\
2010-07-06 quantiles & $5.76$ & $6.05$ & $6.26$ & $0.232$ & $0.304$ & $0.399$ & $77.0$ & $85.2$ & $89.7$ & $7.54$ & $8.27$ & $11.2$ & $0.349$ & $0.667$ & $1.23$ & \\
% cut out 10k/27k
\hline
% 2010-12-25 max. $\log\mathcal{L}$* & \multicolumn{3}{c|}{$793231$} & \multicolumn{3}{c|}{$3.06$} & \multicolumn{3}{c|}{$28.3$} & \multicolumn{3}{c|}{$50.35$} & \multicolumn{3}{c|}{$3.11$} \\
% 2010-12-25 quantiles & $765637$ & $791533$ & $836450$ & $2.80$ & $2.99$ & $3.20$ & $10.55$ & $25.22$ & $39.69$ & $25.09$ & $48.59$ & $65.82$ & $1.86$ & $3.03$ & $4.68$ \\
2010-12-25 max. $\log\mathcal{L}$* & \multicolumn{3}{c|}{$7.93$} & \multicolumn{3}{c|}{$3.06$} & \multicolumn{3}{c|}{$28.3$} & \multicolumn{3}{c|}{$50.3$} & \multicolumn{3}{c|}{$3.11$} & Heterogeneous \\
2010-12-25 quantiles* & $7.66$ & $7.92$ & $8.36$ & $2.80$ & $2.99$ & $3.20$ & $10.6$ & $25.2$ & $39.7$ & $25.1$ & $48.6$ & $65.8$ & $1.86$ & $3.03$ & $4.68$ & \\
% cut out 10k/~34k
\hline
% 2013-02-15 max. $\log\mathcal{L}$ & \multicolumn{3}{c|}{$100.54$} & \multicolumn{3}{c|}{$2.83$} & \multicolumn{3}{c|}{$88.71$} & \multicolumn{3}{c|}{$6.98$} & \multicolumn{3}{c|}{$0.017$} \\
% 2013-02-15 quantiles & $100.52$ & $100.58$ & $100.93$ & $2.83$ & $2.83$ & $2.83$ & $81.42$ & $87.77$ & $88.81$ & $6.94$ & $7.00$ & $11.96$ & $0.015$ & $0.032$ & $0.14$ \\
2013-02-15 max. $\log\mathcal{L}$ & \multicolumn{3}{c|}{$112$} & \multicolumn{3}{c|}{$2.51$} & \multicolumn{3}{c|}{$35.6$} & \multicolumn{3}{c|}{$10.0$} & \multicolumn{3}{c|}{$0.0374$} & Heterogeneous \\
2013-02-15 quantiles & $111$ & $113$ & $119$ & $2.18$ & $2.51$ & $2.52$ & $34.5$ & $49.5$ & $91.4$ & $7.61$ & $9.63$ & $14.3$ & $0.0287$ & $0.0686$ & $0.534$ & \\
% cut out 30k/~100k
\hline
% 2013-04-30 max. $\log\mathcal{L}$ & \multicolumn{3}{c|}{$782536$} & \multicolumn{3}{c|}{$4.56$} & \multicolumn{3}{c|}{$86.05$} & \multicolumn{3}{c|}{$9.15$} & \multicolumn{3}{c|}{$0.25$} \\
% 2013-04-30 quantiles & $741094$ & $770835$ & $795955$ & $4.45$ & $4.55$ & $4.56$ & $77.01$ & $83.84$ & $89.31$ & $8.63$ & $9.39$ & $10.23$ & $0.20$ & $0.27$ & $0.55$ \\
2013-04-30 max. $\log\mathcal{L}$ & \multicolumn{3}{c|}{$7.83$} & \multicolumn{3}{c|}{$4.56$} & \multicolumn{3}{c|}{$86.1$} & \multicolumn{3}{c|}{$9.15$} & \multicolumn{3}{c|}{$0.254$} & Strong Homogeneous \\
2013-04-30 quantiles & $7.41$ & $7.71$ & $7.96$ & $4.45$ & $4.55$ & $4.56$ & $77.0$ & $83.8$ & $89.3$ & $8.63$ & $9.39$ & $10.2$ & $0.201$ & $0.273$ & $0.549$ & \\
% cut out 10k/~22k
\hline
% 2016-02-06 max. $\log\mathcal{L}$ & \multicolumn{3}{c|}{$535204$} & \multicolumn{3}{c|}{$1.36$} & \multicolumn{3}{c|}{$31.73$} & \multicolumn{3}{c|}{$3.59$} & \multicolumn{3}{c|}{$0.037$} \\
% 2016-02-06 quantiles & $526647$ & $543591$ & $562948$ & $1.32$ & $1.36$ & $1.40$ & $27.21$ & $32.13$ & $38.29$ & $3.42$ & $3.55$ & $3.76$ & $0.025$ & $0.052$ & $0.48$ \\
2016-02-06 max. $\log\mathcal{L}$ & \multicolumn{3}{c|}{$5.35$} & \multicolumn{3}{c|}{$1.36$} & \multicolumn{3}{c|}{$31.7$} & \multicolumn{3}{c|}{$3.59$} & \multicolumn{3}{c|}{$0.0365$} & Heterogeneous \\
2016-02-06 quantiles & $5.27$ & $5.44$ & $5.63$ & $1.32$ & $1.36$ & $1.40$ & $27.2$ & $32.1$ & $38.3$ & $3.42$ & $3.55$ & $3.76$ & $0.0251$ & $0.0523$ & $0.481$ & \\
% cut out 10k/~29k
\hline
% 2018-12-18 max. $\log\mathcal{L}$* & \multicolumn{3}{c|}{$637673$} & \multicolumn{3}{c|}{$9.41$} & \multicolumn{3}{c|}{$54.17$} & \multicolumn{3}{c|}{$76.52$} & \multicolumn{3}{c|}{$1.36$} \\
% 2018-12-18 quantiles & $622040$ & $649500$ & $1005178$ & $7.92$ & $9.41$ & $9.43$ & $11.95$ & $43.97$ & $54.25$ & $64.96$ & $104.05$ & $242.59$ & $1.16$ & $2.78$ & $16.87$ \\
2018-12-18 max. $\log\mathcal{L}$* & \multicolumn{3}{c|}{$6.38$} & \multicolumn{3}{c|}{$9.41$} & \multicolumn{3}{c|}{$54.2$} & \multicolumn{3}{c|}{$76.5$} & \multicolumn{3}{c|}{$1.36$} & Strong Aggregate \\
2018-12-18 quantiles* & $6.22$ & $6.50$ & $10.0$ & $7.92$ & $9.41$ & $9.43$ & $12.0$ & $44.0$ & $54.3$ & $65.0$ & $104$ & $243$ & $1.16$ & $2.78$ & $16.9$ & \\
% cut out 2000/~28k
\hline
% 2020-12-22 max. $\log\mathcal{L}$ & \multicolumn{3}{c|}{$848348$} & \multicolumn{3}{c|}{$1.07$} & \multicolumn{3}{c|}{$61.44$} & \multicolumn{3}{c|}{$1.94$} & \multicolumn{3}{c|}{$1.61$} \\
% 2020-12-22 quantiles & $777576$ & $849891$ & $1004013$ & $1.05$ & $1.07$ & $1.10$ & $41.19$ & $62.10$ & $73.91$ & $1.31$ & $1.61$ & $2.69$ & $0.11$ & $0.69$ & $9.59$ \\
2020-12-22 max. $\log\mathcal{L}$ & \multicolumn{3}{c|}{$8.41$} & \multicolumn{3}{c|}{$1.07$} & \multicolumn{3}{c|}{$62.2$} & \multicolumn{3}{c|}{$1.44$} & \multicolumn{3}{c|}{$0.116$} & Weak Homogeneous \\
2020-12-22 quantiles & $7.78$ & $8.50$ & $10.0$ & $1.05$ & $1.07$ & $1.10$ & $41.2$ & $62.1$ & $73.9$ & $1.31$ & $1.61$ & $2.69$ & $0.113$ & $0.693$ & $9.599$ & \\
% cut out 2000/~29k
\hline
\end{tabular}
}
}
\caption{The maximum log-likelihood fit and nested sampling posterior quantiles for the physical parameters of $13$ decameter impactors. Entries marked with an asterisk (*) have had their USG sensor-recorded light curves truncated due to contamination by reflected sunlight from the dust trail at low altitudes. The structural class of each object identified in Section \ref{sec:structural_properties} is also listed.}
% \caption{Comparison of Maximum Log-Likelihood and Nested Sampling Posterior Quantiles for Derived Physical Parameters to Previous Estimates}
% \textbf{Note:} Entries marked with an asterisk (*) have had their USG sensor-recorded light curves truncated due to contamination by reflected sunlight from the dust trail at low altitudes.
\label{tab:decameter}
\end{table*}

From these results, we identify three broad classes of objects within the decameter impactor population. We also correlate the probable escape regions in the main-belt for decameter impactors with these strength classes to examine if their are any obvious links between their dynamical origins and material/structural properties.

\subsection{Structural Properties}\label{sec:structural_properties}

The only material strength classification of large meteoroids and small asteroids proposed to date is given by \citeA{borovicka_two_2020}, which we adopt in this work and briefly summarize here.

In their proposed classification, the strongest objects are composed of Category A material which has tensile strength $20-40$ MPa and represent compact, monolithic objects that can survive passage through the Earth's atmosphere and be recovered as meteorites. 
Category B material has tensile strength $1-5$ MPa and represents macroscopically cracked material weakened by asteroid collisions in space. Favorable entry conditions may allow meteorites to survive to the ground from these objects.
Category C material has tensile strength $0.04-0.12$ MPa and represents collisional debris that has been reassembled and cemented onto the surface of the meteoroid. This grade of material is quickly separated from the meteoroid during atmospheric entry and released at high altitudes.
Finally, they propose the weakest material as ``Category D" type. This material has virtually no tensile strength ($\sim0$ MPa) and represents rubble-pile asteroids loosely bound by gravity or van der Waals forces. This type of material fragments too quickly upon atmospheric entry to be detectable as part of meteoroid ablation and remains speculative \cite{borovicka_are_2015}, though dedicated observations of fireballs early in flight have found formation of wake indicating grain release at very low dynamic pressures \cite<$\sim5$ kPa;>[]{shrbeny_fireball_2020}.

From our analysis, we find that the decameter impactor population shows significant diversity cross-cutting these strength categories even for our small dataset.
The clearest grouping consists of the 14 January 1999, 3 September 2004, 7 October 2004, 6 July 2010 and 22 December 2020 bolides. These are all globally weak objects where in all cases, most ($\gtrsim80\%$) of the object's initial mass was released at relatively low dynamic pressure ($\lesssim1.5$ MPa) and only a small amount ($<1\%$) of the object's initial mass survived to peak dynamic pressure. We suggest that the October 2004 and December 2020 bolides are comprised of weak, heavily fractured category B material released mostly at the main fragmentation point. The January 1999, September 2004 and July 2010 bolides are even weaker objects comprised mostly of category C material, consisting of reassembled collisional debris cemented together with a global tensile strength of $\sim0.1$ MPa, in addition to a small amount of slightly stronger category B material. We therefore suggest these five objects form a relatively distinct class comparable to the Romanian bolide, representing weak, structurally homogeneous (which we term weak homogeneous) bodies that completely disintegrate in the atmosphere and do not produce meteorites \cite{borovicka_january_2017}.

% each releasing $\sim85\%$ of their overall mass in the main fragmentation at $\sim2.8$ and $\sim4.5$ MPa respectively. 
The 30 April 2013 bolide displays similar structural homogeneity to the weak bolides but is significantly stronger overall, releasing $\sim85\%$ of its initial mass in the main fragmentation at $\sim4.5$ MPa. 
Based on the fragmentation behaviour, we suggest the April 2013 bolide is structurally similar to (though somewhat weaker than) the L5 chondrite Park Forest \cite{simon_fall_2004}, which we estimate as having released $\sim85\%$ of its mass at the $\sim6.3$ MPa main burst and whose fit light curve is very similar in shape
(compare Figure \ref{fig:park_forest_lc}a) and Figure S$17$ in Supporting Information S$1$). It forms a class of its own which we term strong homogeneous.
% lc before posteriors
% In this regard we suggest it is structurally similar to (though somewhat weaker than) the Park Forest fireball which released $\sim85\%$ of its mass in the main burst at $7$ MPa \noteic{(should we use the \citeA{brown_orbit_2004} numbers or ours for comparison here?)} and reached a peak dynamic pressure of $14$ MPa; indeed, the fit light curves are very similar in shape.

In contrast, the 1 February 1994, 9 December 2006, 25 December 2010 and 6 February 2016 bolides display significant structural heterogeneity, characterized by multiple, extended fragmentation episodes. In all cases at least five major fragmentations were identified starting at $\sim1$ MPa, and with no single fragment comprising more than $\sim40\%$ of the object's total mass. The maximum strength of these objects varies, with the last fragmentation at $\sim35$ MPa, $\sim5.5$ MPa, $\sim8$ MPa, and $\sim3.5$ MPa respectively. 
% We also note that the dynamic pressure at low altitude for the December 2010 bolide is likely overestimated since the light curve has been truncated due to contamination from reflected sunlight.
Based on the dynamic pressure range and numerous large fragmentations, we suggest that the latter three objects are composed almost entirely of category B material, with tensile strength ranging from $1-5$ MPa. The February 1994 bolide stands out an an exceptionally strong object that likely contained a significant amount of extremely strong category A material or as suggested previously was perhaps made of iron \cite{nemtchinov_assessment_1994}.
We therefore suggest these four bolides represent a class of structurally heterogeneous objects that incrementally fragment along cracks when the dynamic pressure reaches the tensile strength of that fragment, often with a small portion of the object (up to a few percent of the initial mass) surviving up to a high peak dynamic pressure. These strong inclusions may represent essentially intact ``boulders" of high strength, similar to that found by \citeA{borovicka_trajectory_2013} for the Chelyabinsk bolide whereby $\sim1\%$ of the initial meteoroid mass was composed of relatively strong meter-sized ``boulders".

The 15 February 2013 Chelyabinsk bolide represents a special case as it is the only decameter impactor for which detailed ablation/fragmentation modeling has previously been conducted and we discuss it here separately. Analysis by \citeA{borovicka_trajectory_2013} using the light curve of \citeA{brown_500-kiloton_2013} in conjunction with ground-based video footage and acoustic analysis suggests that the first major fragmentation occurred at a height of $45$ km and dynamic pressure of $\sim0.7$ MPa, followed by a series of $11$ individual fragmentations between $40-30$ km height and $1-5$ MPa dynamic pressure in which most of the total mass was released. A tiny portion ($\sim1\%$) penetrated much deeper into the atmosphere as a collection of large, strong fragments, surviving dynamic pressures of $10-20$ MPa before fragmenting. At least one fragment impacted into Lake Chebarkul and was later recovered as a $570$ kg meteorite. 
% after which $<1\%$ of the initial mass remained.
To simplify our model, we chose to approximate the series of $11$ fragmentations between $40-30$ km with three large fragmentation episodes as this produces a reasonable fit to the observed light curve. The light curve is the only data constraint we use here for consistency in our modeling approach with other USG bolides. However, based on the large number of fragmentations and dynamic pressure range identified by \citeA{borovicka_trajectory_2013} using additional ground-based video data we suggest that Chelyabinsk is materially similar to our structurally heterogeneous object class (which we term simply heterogeneous as they span the spectrum from weak to strong) and composed mostly of category B material with a global strength of order $1-5$ MPa.

The 8 October 2009 and 18 December 2018 fireballs are relatively unique objects that we discuss separately. Both are very energetic ($30-50$ kT TNT), occurred during daylight and based on their light curves remained mostly intact until the main fragmentation at $9-10$ MPa. This is well above the maximum strength of cracked category B material ($\sim5$ MPa) yet much less than the tensile strength of pristine category A meteorites ($\sim20-40$ MPa).

We suggest that the peak dynamic pressure for these objects is significantly overestimated as light curves for these two objects have been truncated at low altitudes (due to likely persistent trail emission following the main airburst) but this has minimal effect on the inferred dynamic pressure at the main fragmentation and mass loss fraction up to the height of peak brightness. Despite the dimmer late persistent emission which makes it appear as though there was very low altitude ablation, most of the energy deposition occurs near the height of peak brightness where dust emission is less significant and hence the energies for these objects are likely not overestimated.
Indeed, analysis by \citeA{silber_infrasonic_2011} of worldwide infrasound from the 8 October 2009 fireball suggests a most likely energy of $50$ kT TNT. More recently, \citeA{arrowsmith_bolide_2021} analyzed infrasound from the 18 December 2018 bolide, finding an average period of $14.8$ sec using select stations with high signal-to-noise ratio. We performed a similar analysis of signals from $15$ infrasound stations and find a mean period of $16.3$ sec.
Using the multi-station average relation of \citeA{ens_infrasound_2012}, these periods correspond to source energies of $35$ and $49$ kT TNT respectively, both of which are close to the USG-reported energy of $49$ kT TNT.

These two bolides plausibly represent a separate, unique class of stronger objects whose structure is dominated by microcracks or microporosity rather than macroscopic fractures created by asteroid collisions alone. We term these objects strong aggregates.

\subsection{Source Regions}

\begin{figure*}
    \centering
    \includegraphics[width=1.\linewidth]{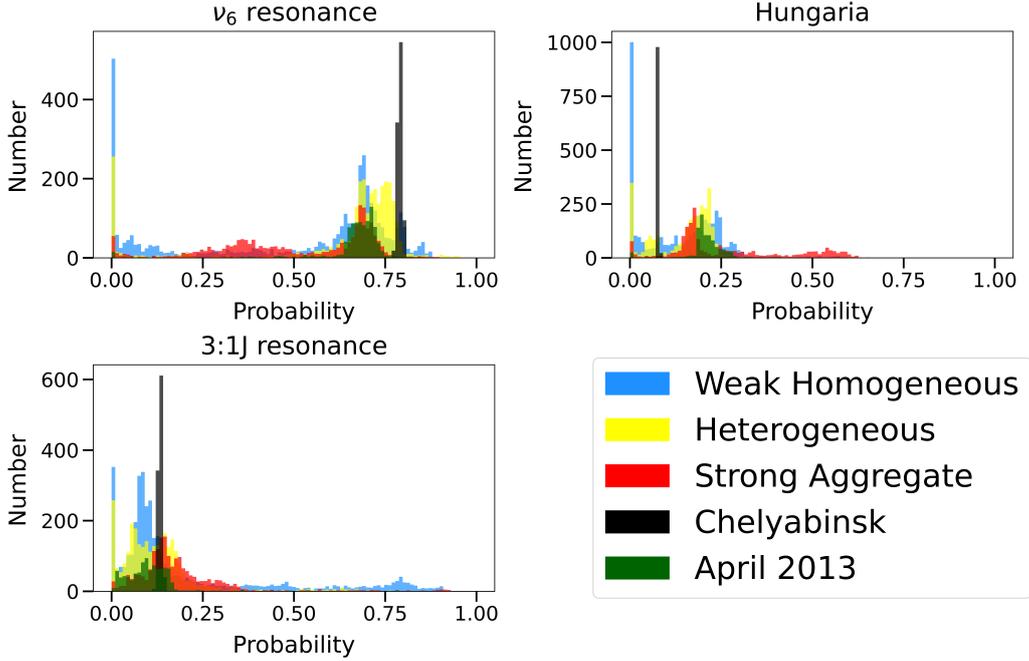}
    \caption{Histograms of source region probabilities for all USG sensor-detected decameter impactors and corresponding Monte Carlo clones for each of the three broad strength categories (weak homogeneous, heterogeneous, and strong aggregate) identified in the text, computed using the NEO model of \citeA{granvik_debiased_2018}. The Chelyabinsk (February 2013) and April 2013 bolides are plotted separately. Only the three primary source regions of decameter impactors (the $\nu_6$ secular resonance, Hungaria family and 3:1 mean-motion resonance with Jupiter) identified by \citeA{chow_decameter-sized_2025} are shown here, as the contributions of other source regions to the decameter impactor population are negligible.}
    \label{fig:escape_regions}
\end{figure*}

We also analyze the origins of these decameter impactors from the main asteroid belt to see if any correlations exist with their structural properties. \citeA{chow_decameter-sized_2025} used the NEO model of \citeA{granvik_debiased_2018} to determine the probable source regions of the decameter impactors from their pre-impact orbits, generating $1000$ Monte Carlo clones per object to estimate orbital uncertainties. The decameter impactor population was found to originate from the $\nu_6$ secular resonance with a probability of $\sim70\%$, the Hungaria group with a probability of $\sim20\%$ and the 3:1 Jupiter mean-motion resonance with a probability of $\sim10\%$. No significant escape probabilities were found from the mid to outer main belt. Figure \ref{fig:escape_regions} shows the decameter impactor source region probabilities computed in \citeA{chow_decameter-sized_2025} broken down by each of the broad strength classes (weak homogeneous, heterogeneous, and strong aggregate) identified in this work. The Chelyabinsk (February 2013) and April 2013 bolides are plotted separately: the former as its orbit (and therefore probable source region) is far better characterized than the other decameter impactors, and the latter as it is significantly stronger than the other homogeneous bodies.

% results broken down by each of the broad strength classes identified in the previous section.
% generated to estimate the orbital uncertainties.
% each of the decameter impactors
% determine the probable source regions for the decameter impactors
% and $1000$ Monte Carlo 
% We have taken
% Figure \ref{fig:escape_regions}
% We also analyze the source regions of impactors in these broad strength classes to determine where each of these subclasses of objects within the decameter impactors originate.
% separated 

To zeroth order, the distribution of source region probabilities for each of our strength classes is generally consistent with the population-level results found in \citeA{chow_decameter-sized_2025}, with no significant difference between expected source regions based on material strength. However, we note that our proposed classification and associated conclusions are tentative as they are limited by small-number statistics. As such, we do not strictly rule out a connection between dynamical origin and material strength for the decameter impactors, but merely conclude that there is no strong evidence for such a connection based on the currently limited number of detections.
% suggesting that there is no obvious connection between dynamical origin and material strength for the decameter impactors.

% significant difference between expected source regions based on b
% the population-level source region distribution of \citeA{chow_decameter-sized_2025} is roughly consistent
% Figure \ref{fig:escape_regions} shows the source region probabilities computed using the model of \citeA{granvik_debiased_2018} for the impactors (and their corresponding Monte Carlo clones) in each of the broad strength classes identified here, as well as the 2013 April and 2013 February (Chelyabinsk) bolides, which are plotted separately. In \citeA{chow_decameter-sized_2025}, we found the decameter impactor population as a whole to originate from the $\nu_6$ secular resonance with a $\sim70\%$ chance, the Hungaria group with a $\sim20\%$ chance and the 3:1 Jupiter mean-motion resonance (MMR) with a $\sim10\%$ chance. 

% These results appear to still hold for each of the subpopulations identified here, suggesting there is no obvious connection between dynamical origin and material strength for decameter impactors.
% The source regions for the decameter impactors are largely agnostic of material strength.

\section{Discussion}\label{sec:discussion}

\subsection{Uncertainties and Limitations in Analysis}\label{sec:uncertainties_limitations}
Here we discuss sources of uncertainty as well as some of the choices made for our light curve modeling procedure and results of alternative methods that were explored. 
While we adopted fixed luminosity uncertainty in our modelling procedure, we also attempted to fit the light curves with intrinsic scatter/prediction error as a free parameter in the model to further improve our posterior estimates. This was done using a modified log-likelihood formalism $\mathcal{L'}$ that adds a free intrinsic scatter term $\sigma_{\mathrm{scat}}^2$ (parameterized as $\log$-$\sigma_{\mathrm{scat}}^2$ in the nested sampling) to the error, in quadrature with the observational uncertainty $\sigma_i^2$, according to
\begin{equation}
    \ln\mathcal{L'}\left(\bm\Theta, \sigma_{\mathrm{scat}}^2\right) = -\sum_{i=1}^{N}\left(\frac{\left(L_i - \bar{L}_i\left(\bm{\Theta}\right)\right)^2}{2\left(\sigma_i^2 + \sigma_{\mathrm{scat}}^2\right)} + \ln\sqrt{2\pi\left(\sigma_i^2+ \sigma_{\mathrm{scat}}^2\right)}\right)\label{eqn:logL_scatter}.
\end{equation}
In this case the first term of Equation \ref{eqn:logL_scatter} corresponds to the traditional reduced $\chi^2$, while the second term acts as an error term that effectively penalizes for large values of scatter. We tested both uniform and inverse-gamma priors (the latter being the conjugate prior of a Gaussian) for the scatter in the nested sampling, with bounds chosen so the prior encompasses the sample variance of the residuals for a manually-fit initial solution. However, in both cases, we found that the nested sampler often performed very poorly for many fireballs, becoming stuck in regions of parameter space that produced solutions completely incongruous with the observed light curves. This could be due to large differences in intensity throughout the duration of the light curves, often up to several orders of magnitude. 
% Indeed, initial efforts to infer intrinsic scatter when applying the method introduced in this paper to model light curves of smaller meteoroids, whose intensities vary by much less, have been generally successful (Vovk et al., in prep).
We also explored fitting the USG light curves using the luminous efficiency model of \citeA{borovicka_two_2020} instead of a constant value based on the CNEOS-reported radiated and kinetic energies for each impactor. However, this produced initial masses that were several orders of magnitude too large for the validation cases, likely as a result of the USG sensor bandpasses being significantly different from ground-based bandpasses. As such, we chose to use a constant luminous efficiency derived from USG light curve comparisons to infrasound energies as reported in \citeA{brown_flux_2002} as it produced mass estimates generally consistent with the literature.

%%% Maybe include this?
% The velocities and radiants recorded by USG sensors, which are generally considered less accurate \cite<e.g.>[]{devillepoix_observation_2019, brown_proposed_2023, chow_decameter-sized_2025} than the energy determinations.

For the validation cases, we considered only the major fragmentation points identified by previous works where light curves were fit manually. This was done in order to facilitate direct comparison with previous studies and reduce the complexity of the nested sampling procedure, as each additional fragment significantly increases the dimensionality of the parameter space.
While several models with varying numbers of fragments could in principle be fit to each light curve and rigorously compared using evidence-based metrics such as the Bayes factor (the ratio of evidences $\mathcal{Z}_1/\mathcal{Z}_2$ for two different models) or the Bayesian information criterion \cite<BIC;>[]{schwarz_estimating_1978}, we have foregone this approach as the USG light curve data is not precise enough for such a procedure to be robust. Instead, we merely note that for the validation cases, the posterior distributions for initial mass, dynamic pressure and mass released at main fragmentation inferred by our models under conservative, flat priors solely from the USG light curves are consistent with previous estimates using independent observations. The resulting fits qualitatively match the observed light curves well, particularly near peak brightness where most of the energy (and mass) is released.

For completeness, we also reanalyzed the light curves of all validation events using the reported meteorite density as the initial, pre-impact bulk density (excluding the Romania fireball as no meteorites were recovered) instead of the $1500$ kg m$^{-3}$ estimate of \citeA{chow_decameter-sized_2025}, with the idea that this represents an upper limit for the bulk density in these cases.
The reported meteorite bulk densities for all validation events are summarized in Table \ref{tab:meteorite_densities}. The only decameter impactor for which meteorites were recovered is Chelyabinsk, which was an LL5 chondrite with meteorite bulk density of $\sim3300$ kg m$^{-3}$ \cite{popova_chelyabinsk_2013-1}.
\begin{table*}[t]
    \begin{center}
    \begin{tabular}{|l|c c c|}
    \hline
    \textbf{Meteorite} & \textbf{Classification} & \textbf{Bulk Density (kg m$^{-3}$)} & \textbf{Reference} \\
    \hline
    Tagish Lake & C$2$-ungrouped & $1670$ & \citeA{flynn_physical_2018} \\
    Mor\'{a}vka & H$5$ & $3590$ & \citeA{flynn_physical_2018} \\
    Park Forest & L$5$ & $3900$ & \citeA{flynn_physical_2018} \\
    Ko\v{s}ice & H$5$ & $3430$ & \citeA{flynn_physical_2018} \\
    Romanian Fireball & & no meteorites recovered & \\
    Sari\c{c}i\c{c}ek & HED & $2929$ & \citeA{unsalan_saricicek_2019} \\
    Flensburg & C$1$-ungrouped & $1984$ & \citeA{borovicka_trajectory_2021} \\
    \hline
    \end{tabular}
    \end{center}
    \caption{The classifications and weighted-average bulk densities of meteorites recovered from the seven validation fireball events considered in this work.}
    \label{tab:meteorite_densities}
\end{table*}
We found that this reanalysis did not significantly change the inferred parameter posterior distributions for any of the validation events, a result consistent with previous studies where fireball light curves were manually fit \cite<e.g.>[]{borovicka_maribo_2019}. % \noteic{others?}.

For the velocity uncertainties, we attempted to empirically ground USG velocity errors using a set of common USG sensor and ground-based fireball measurements, as described in \citeA{chow_decameter-sized_2025}. The majority of fireballs in this calibration dataset had speed estimates that agreed to within $2$ km s$^{-1}$ between ground-based and USG values, though a small number had differences approaching $8$ km s$^{-1}$. Given that these simultaneously ground-based and USG observed events were of lower energy and shorter duration than our larger decameter impactors, we expect our uncertainties to be lower. Taking a typical speed uncertainty of $2$ km s$^{-1}$, this implies an uncertainty of $\sim20-30\%$ in absolute dynamic pressure, though the relative difference in dynamic pressure between fragmentation points can be expected to be much less.
The same calibration dataset yields a mean height difference for peak brightness between USG and ground-based measurements of $3$ km \cite{brown_proposed_2023}. This is less than half an atmospheric scale height and therefore the height uncertainties should produce less than a factor of two uncertainty in absolute dynamic pressure. 

Finally, we note that as the sensitivity of the USG sensors is near magnitude $-16.5$, the brightness in the earliest portion of flight is not recorded. Measurements in this early luminous phase often show that distinct fragmentation in the form of wake formation does occur at high altitudes under very low dynamic pressure \cite{shrbeny_fireball_2020}. We are entirely insensitive to this phase of fireball flight other than to say that such high altitude mass loss does not appear to be globally significant for decameter impactors. 

\subsection{Physical Interpretation and Implications for Planetary Defence}\label{sec:planetary_defence}
% It is commonly assumed that strength (and hence fragmentation behaviour) for meter-sized and larger meteoroids follows a Weibull-like power-law \cite{popova_very_2011}. 
It is commonly assumed that strength (and hence fragmentation behaviour) for meter-sized and larger meteoroids follows a Weibull-like power-law \cite<e.g.>[]{popova_very_2011, flynn_physical_2018}.
This power-law is grounded in observations of terrestrial rock fracturing which show that larger rocks are weaker than smaller rocks on average \cite{hartmann_terrestrial_1969}. In terms of the change in fragment strength with mass, this relationship follows \cite{popova_very_2011}
\begin{equation}
    \sigma = \sigma_s \left(M_{s}/M\right)^\alpha, \label{eqn:weibull}
\end{equation}
where $\sigma$ is the strength of a fragment of mass $M$, $\sigma_s$ is the strength of a smaller reference mass of mass $M_s$ and $\alpha$ is a power-law coefficient relating the change in strength across mass scales. 
% This is an extreme idealization as real fireballs follow a wide range of apparent strength scaling with mass.
We note that while the tensile strength of individual fragments on average increases at smaller sizes, the Weibull distribution is an idealization; real fireballs display a wide variation in strength scaling depending on the object and the meteorite type. % \cite{ostrowski_physical_2019}.
Nevertheless, understanding bounds for $\alpha$ is important for planetary defense, where broad statistical distributions are used prior to impact to bracket expected ground damage \cite<e.g.>[]{wheeler_fragment-cloud_2017, mathias_probabilistic_2017, wheeler_atmospheric_2018}.
Past estimates of $\alpha$ have generally ranged from $0.1-0.5$ \cite{svetsov_disintegration_1995, popova_modelling_2019} though recognition of the large range found in fireball data have led those focusing on impact effects related to planetary defense to broaden estimates to as much as $0.05-1$ \cite{wheeler_fragment-cloud_2017} when considering the effect on fireball energy deposition.

\begin{figure*}
\centering
\includegraphics[width=1.\linewidth]{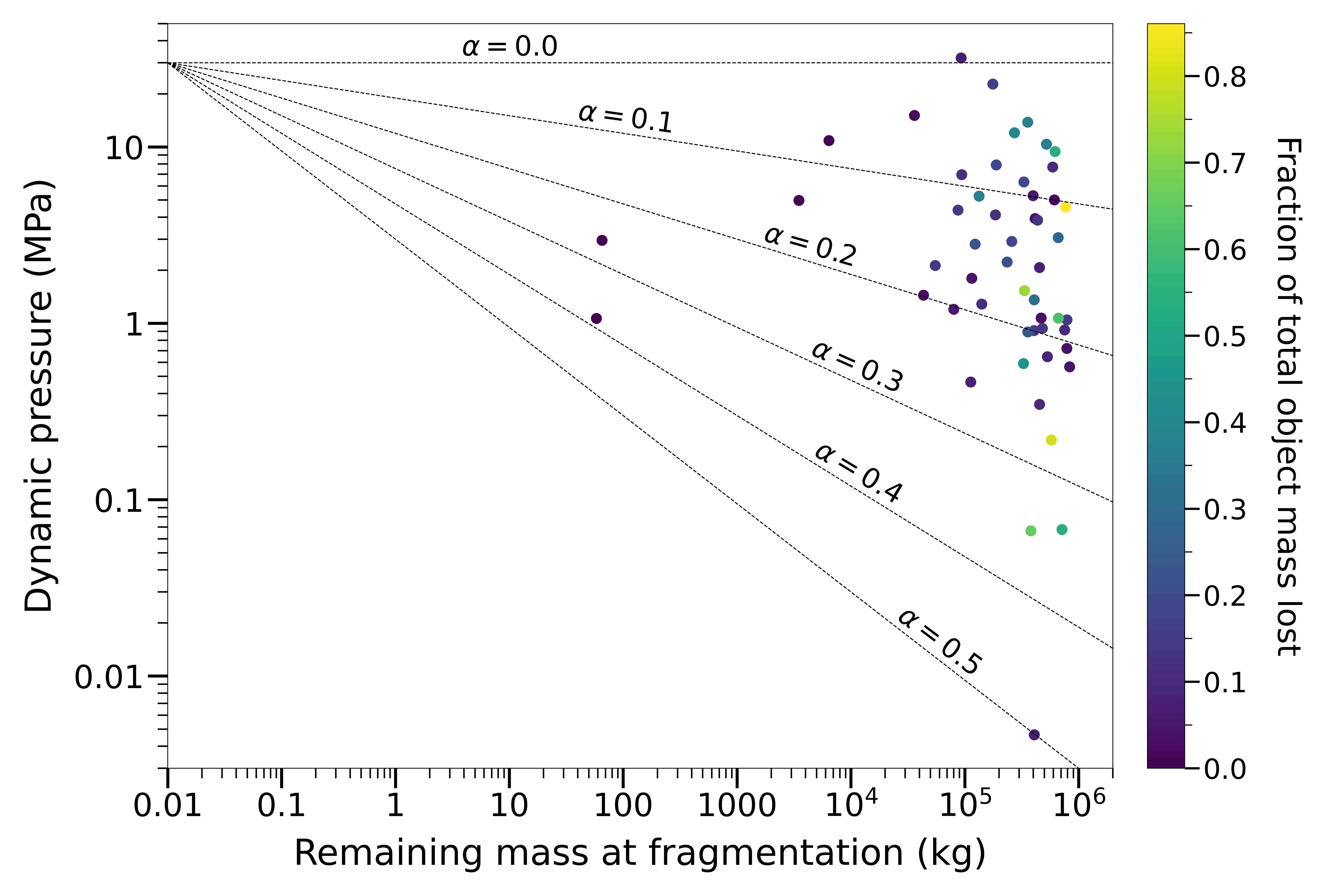}
\caption{
Dynamic pressure plotted against mass remaining at each fragmentation point for the maximum log-likelihood solutions of all $13$ decameter impactors, with each point coloured by the fraction of total mass released. Assuming a tensile strength of $30$ MPa for a $0.01$ kg fragment, the black dashed lines show Weibull power-laws given by Equation \ref{eqn:weibull} for values of the strength scaling parameter $\alpha$ ranging from $0.1$ to $0.5$. We find that values of $\alpha$ ranging from $\sim0-0.3$ are appropriate to describe the fragmentation behaviour of decameter-size objects.
% The red solid line and shaded region show the empirical best fit to the data and the $1\sigma$ error in the estimated value of $\alpha$. % sing least-squares optimization and the $1\sigma$ error.
}
\label{fig:weibull_plots}
\end{figure*}

In Figure \ref{fig:weibull_plots} we show the range of fragmentation strengths as a function of mass estimated for our decameter impactors. Here we adopt a $30$ MPa tensile strength for a $0.01$ kg fragment for consistency with previous work \cite{popova_very_2011}.
We find a wide spread in the strength scaling parameter $\alpha$, with the commonly adopted range of $0.1 - 0.5$ well supported by our measurements to decameter sizes. 
For comparison, \citeA{wheeler_fragment-cloud_2017} found that values of $\alpha$ between $0.1 - 0.3$ were best able to replicate typical burst heights when estimating equivalent point-source burst altitudes.
Fitting equation \ref{eqn:weibull} to the observed data using least-squares optimization yields an empirical best-fit value of $\alpha \approx 0.12 \pm 0.011$, though most of the observed data deviates from a theoretical Weibull power law, as shown in Figure \ref{fig:weibull_plots}. Our goal here is merely to produce an empirical estimation of the strength and fragmentation behaviour of decameter-size objects, which previous planetary defense studies have identified as the most likely to cause ground damage in the near future \cite<e.g.>[]{boslough_updated_2015}, for the first time.
As such, we simply state that the majority of decameter impactors are fit well qualitatively for values of $\alpha$ ranging from $\sim0 - 0.3$. 

An important new result from fragmentation measurements of decimeter- to meter-sized chondritic meteoroids by \citeA{borovicka_two_2020} was the first evidence for two distinct fragmentation stages. The first stage was found to consist of fragmentation releasing mass under dynamic pressures of $0.04-0.12$ MPa. \citeA{borovicka_two_2020} interpret this phase as evidence for weakly cemented or aggregate fines (i.e. category C material) held together after having been potentially produced during impact on the original (larger) parent body. Of the $21$ chondritic-like fireballs/meteorite-dropping bolides examined, this phase released $40\%$ or more of the total mass in the majority of cases.  

Few events studied by \citeA{borovicka_two_2020} showed significant fragmentation between $0.12$ and $0.9$ MPa -- material bound at these strengths is surprisingly absent in chondritic-like fireballs. They noted that this second phase of fragmentation was found to last from $0.9 - 5$ MPa and typically released less total mass than the first stage as well as occasionally being absent for some bolides. This phase was interpreted as being due to more competent rocky material weakened by collisional cracks (category B material), reducing their strength compared to meteorites found in the lab by an order of magnitude or more.  

\begin{figure*}
\centering
\includegraphics[width=1.\linewidth]{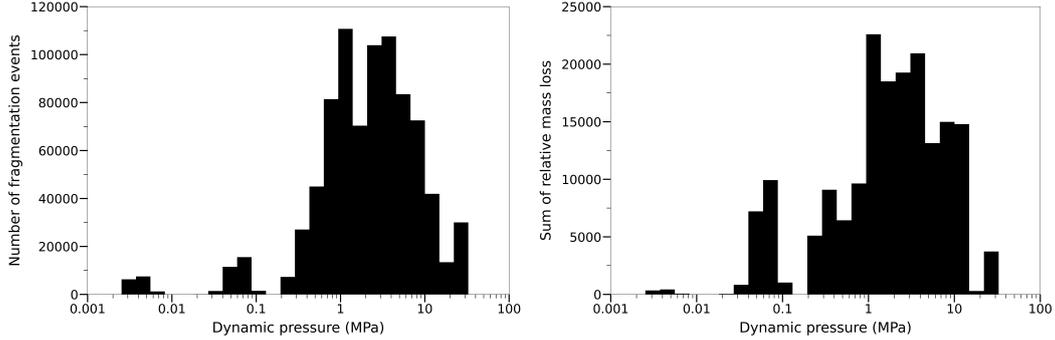}
    \caption{
    Histograms showing the distribution of dynamic pressures at major fragmentation points for all decameter impactors, using $15,000$ samples drawn randomly from the nested sampling posteriors of each impactor. The left histogram is unweighted, while the right histogram shows each fragmentation event weighted by the relative mass loss.
    }
    \label{fig:dyn_pres_histograms}
\end{figure*}

Similar to \citeA{borovicka_two_2020}, we find here that the majority ($10$ of $13$) of our decameter impactors begin fragmenting under dynamic pressures less than $1$ MPa. As shown in Figure \ref{fig:dyn_pres_histograms}, there is also evidence for two stages of fragmentation (compare to Figure 6 of \citeA{borovicka_two_2020}): one with a peak in mass loss at $\sim0.04-0.09$ MPa and a later peak occurring near $1-4$ MPa. In contrast to the smaller objects studied by \citeA{borovicka_two_2020}, we find the second stage of fragmentation is where the majority of mass is released for decameter objects while comparatively little mass is released in the first stage. While our peaks are distinguishable and compare well with the stages identified by \citeA{borovicka_two_2020}, our ranges of fragmentation are more spread out. We suggest this may be a consequence of the lower accuracy of the heights/light curves from USG data and the limited number of fragmentation points adopted in our modeling approach compared to the higher quality data and more detailed modeling employed by \citeA{borovicka_two_2020}. 

We do not have sufficient information to determine whether our objects are chondrite-like material, carbonacous chondrites or other meteorite analogs. We note that \citeA{borovicka_two_2020} explicitly mention the apparently different fragmentation behaviour of the fireball producing the Maribo carbonaceous chondrite fall \cite{borovicka_maribo_2019} and another EN-observed fireball for which similarly high-quality flight data was available. %%%% after
These objects seemingly do not show the two stage fragmentation behaviour, but rather have a large series of smaller fragmentations ranging from $0.25 - 4.3$ MPa and $0.28 - 1.37$ MPa respectively. 
%%% NEW PARAGRAPH HERE
These ranges are very similar to the $0.44 - 1.7$ MPa tensile strength found for meter-size boulders on the asteroid Bennu \cite{ballouz_bennus_2020} and the $0.2-0.28$ MPa tensile strength estimate for a similar boulder on the asteroid Ryugu inferred from thermal measurements \cite{grott_low_2019}, yet less than the tensile strength for a variety of meteorites recovered on Earth \cite<\textgreater10 MPa, except for C2-ungrouped, CI and CM chondrites;>[]{ostrowski_physical_2019} as well as recovered Ryugu samples of $1$--$8$ mm size \cite{nakamura_formation_2022}.
As noted earlier, this reflects the overall increase in strength at smaller sizes.
% We note that while the tensile strength of individual fragments on average increases at smaller sizes, the Weibull distribution is an idealization; real fireballs display a wide variation in strength scaling depending on the object and the meteorite type \cite{ostrowski_physical_2019}.
Among our decameter impactor sample, examination of Figure \ref{fig:mass_cdfs} shows that the 6 February 2016 and 22 December 2020 bolides most closely resemble the boulders, each having half a dozen discrete fragmentation events between $\sim0.5-3$ MPa.

That our decameter objects release more of their mass from cracked, rocky fragments may simply be a consequence of scale. The smaller decimeter- to meter-sized fireballs examined by \citeA{borovicka_two_2020} might be expected to release more cemented/weakly bound material from near-surface regions first, as these are proportionally a larger fraction of the volume for small objects. For our larger fireballs, the cemented material (if in an outer layer) may make up a much smaller total fraction of the object's mass. We propose that at large (decameter) sizes, most impactors are composed of larger blocks riddled with cracks that fail at strengths between $\sim0.3 - 10$ MPa.  

% \begin{figure*}
% \centering
% \begin{minipage}{0.49\textwidth}
%   \centering
%   \includegraphics[width=1.\linewidth]{dyn_pres_hist.png}
% \end{minipage}
% %
% \begin{minipage}{0.49\textwidth}
%   \centering
%   \includegraphics[width=1.\linewidth]{dyn_pres_hist_weighted.png}
% \end{minipage}
%     \caption{
%     Histograms showing the distribution of dynamic pressures at major fragmentation points for all decameter impactors, using $15,000$ samples drawn randomly from the nested sampling posteriors of each impactor. The left histogram is unweighted, while the right histogram shows each fragmentation event weighted by the relative mass loss.
%     }
%     \label{fig:dyn_pres_histograms}
% \end{figure*}

% \noteic{For luminous efficiency we tried using the model of \citeA{borovicka_two_2020} to compute luminous efficiency but it gave mass values that were far too high (many orders of magnitude) for all the validation cases}

% \noteic{We also tried modelling the objects using reported meteorite densities as bulk density for the validation cases as an upper limit for bulk density but this generally did not affect the results}

\section{Conclusions}\label{sec:conclusions}

%%%% MAIN
In this work we have presented a new Bayesian inference method for robust inference of physical parameters of meteoroids. We fit a light curve to observations using the semi-empirical fragmentation model of \citeA{borovicka_kosice_2013} and estimate posterior distributions of the model parameters using dynamic nested sampling. We then validate our method by applying it to the newly released USG light curves of seven fireball events also observed separately by ground-based instruments.
Finally, we apply this method to the USG light curves of $13$ decameter impactors identified in \citeA{chow_decameter-sized_2025} to conduct the first population-level analysis of their material strength and physical structure.
We summarize our major conclusions as follows:
\begin{enumerate}
    \item The parameter values inferred by nested sampling when starting from wide, relatively uninformative priors and using only the USG sensor light curves are generally consistent with previous estimates determined by manual entry modelling of ground-based observations. The quantities most consistent with previous estimates are the dynamic pressure and mass released at the main fragmentation point. The initial mass is also generally consistent when taking into account systematic uncertainties in light curve calibration. The peak dynamic pressure can often be overestimated, especially in cases where an extended dust trail is present at low altitudes.
    % our nested sampling-based method starting from wide, relatively uninformative priors using only the USG sensor light curves
    % are generally consistent with those estimates determined via independent ground-based observations.
    % are consistent with the inferred parameter values from our nested sampling-based method starting from wide, relatively uninformative priors using only the USG sensor light curves. 
    % starting from wide, relatively uninformative priors 
    % determined via other methods. We find that the quantities most consistent with previous estimates are the dynamic pressure and mass released at the main fragmentation point. 
    % The peak dynamic pressure can be overestimated, especially in cases where an extended dust trail is present at low altitudes.
    % The least accurate is the peak dynamic pressure, which can be overestimated if the 
    % \item Validated our method that uses ONLY USG against ground-based observations; initial mass, dynamic pressure and mass released at main fragmentation are consistent with previous estimates, peak dynamic pressure and mass at peak dynamic pressure can be overestimated especially if there is an extended dust trail like with Moravka, Romania and Flensburg (Moravka/Flensburg fell in the daytime)
    \item The decameter impactor population is structurally diverse within even the small number of known impactors. We identify the following three broad structurally distinct classes, based on the material strength classification proposed by \citeA{borovicka_two_2020}.
    \begin{enumerate}
        \item \textit{Homogeneous meteoroids}. This group consists mainly of weak homogeneous objects such as the 14 January 1999, 3 September 2004, 7 October 2004, 6 July 2010 and 22 December 2020 fireballs. Most ($>80\%$) of the object's mass is released below $1.5$ MPa, often in just one or two large fragments, with less than $1\%$ of the mass surviving to peak dynamic pressure. These objects are comprised of weak category B and some category C material. The 30 April 2013 bolide stands out in this broad group as noticeably stronger overall and might be better categorized as a strong homogeneous meteoroid. 
        \item \textit{Heterogeneous meteoroids}. This group consists of the 1 February 1994, 9 December 2006, 25 December 2010, 15 February 2013 and 6 February 2016 fireballs. The object undergoes numerous large fragmentations starting at $\sim1$ MPa, with no single fragment making up more than $\sim40\%$ of the object's initial mass. These objects are comprised of category B material, and in the case of the 1 February 1994 fireball a significant amount of category A material as well.
        \item \textit{Globally strong meteoroids - Strong Aggregates}. This group consists of the 8 October 2009 and 18 December 2018 fireballs. Most ($>80\%$) of the object's mass is released at the main fragmentation at $9-10$ MPa. These bodies could possibly represent a unique class of objects whose global strength is between that of category B and A material, whose structure is primarily dominated by microporosity rather than large cracks.
    \end{enumerate}
    We do not find any significant difference in the global strength of most decameter impactors compared to the smaller meteoroids examined by \citeA{borovicka_two_2020} and earlier by \citeA{popova_very_2011}. This is consistent with the notion that from centimeter-decameter sizes the strength of meteoroids is determined mainly by their unique collisional history as opposed to intrinsic material properties.
    \item The probable source regions of each of the classes of decameter impactors identified above qualitatively match those of the population as a whole found by \citeA{chow_decameter-sized_2025}, with $\sim70\%$ of objects coming from the $\nu_6$ secular resonance, $\sim20\%$ from the Hungaria family and $\sim10\%$ from the 3:1 mean-motion resonance with Jupiter. There is no evidence for any link between the physical structure/material strength of decameter impactors and their dynamical origin.
    \item We identify two distinct fragmentation phases for the decameter impactor population in which most mass is lost, with the first phase occurring at $\sim0.04-0.09$ MPa and the second phase at $\sim1-4$ MPa. % , similar to the results of \citeA{borovicka_two_2020} for meter-size fireballs. 
    This two-stage model of fragmentation and the corresponding dynamic pressures are very similar to those found by \citeA{borovicka_two_2020} for decimeter- to meter-size fireballs.
    % similar to the findings of \citeA{borovicka_two_2020} for meter-size fireballs (where 
    % The two-stage model and fragmentation pressures are very similar to the findings of \citeA{borovicka_two_2020} for meter-size fireballs.
    % from $\sim0.04-0.09$ MPa for the decameter impactor population. 
    % However, the second phase, corresponding to category B material, dominates mass loss for decameter-size objects. 
    % Similarly to the findings of \citeA{borovicka_two_2020} for meter-size fireballs. 
    However, while the first phase (representing category C material) dominates mass loss for meter-size objects, we find that the second phase (representing category B material) is instead dominant for larger decameter-size objects. % For decameter-size impactors, most mass is released beginning around $\sim1$ MPa and can continue up to $\sim10$ MPa.
    \item The main implication of our strength study of decameter-size bolides is in the fragmentation parameters to be used in planetary defense models for similarly sized impactors. 
    In particular, when rock strengths are modelled using a Weibull power law, the main mass release stage of such large impactors is best captured for values of the strength scaling parameter $\alpha$ between $\sim0-0.3$. Moreover, ensuring the main fragmentation stage resides between $\sim1-10$ MPa best captures the observed fragmentation process for decameter impactors.
    % In particular, for treatments using the Weibull fragmentation approach, $\alpha$ values between $\sim0 - 0.3$ may best capture the main mass release stage of such large objects. Moreover, ensuring the main fragmentation stage resides between $1 - 10$ MPa appears to best capture the observed fragmentation process for decameter impactors.
\end{enumerate}
The nested sampling method represents a powerful tool to estimate the physical properties of any meteoroid with an observed light curve, and can naturally be extended to fit for meteoroid dynamics as well when such measurements are available. 
We anticipate future applications of this method to other fireball light curves (including the hundreds of meter-size objects detected by USG sensors) in order to better understand the physical properties of meteoroids.

\section*{Conflict of Interest}

The authors declare there are no conflicts of interest for this manuscript.

\section*{Open Research Section}

All USG sensor fireball data, including light curves, are publicly available on the CNEOS website at \url{https://cneos.jpl.nasa.gov/fireballs/}. The semi-empirical model of \citeA{borovicka_kosice_2013} is implemented using the \texttt{MetSim} software \cite{vida_direct_2023}. The dynamic nested sampling was conducted with the \texttt{dynesty} software package \cite{speagle_dynesty_2020, sergey_koposov_joshspeagledynesty_2024} in Python.

\acknowledgments
We sincerely thank the anonymous reviewers for providing feedback on an earlier version of this manuscript and Josh Speagle for providing many helpful discussions about \texttt{dynesty}. Funding for this work was provided in part by the Meteoroid Environment Office of NASA through co-operative agreement 80NSSC
24M0060, the Natural Sciences and Engineering Research Council of Canada and the Canada Research Chairs program. 
% The authors declare that they have no known conflicts of interest that are not apparent from affiliations or funding.

%%%%%%%%%%%%%%%%%%%%%%%%%%%%%%%%%%%%%%%%%%%%%%%
% REFERENCES and BIBLIOGRAPHY
%
% \bibliography{<name of your .bib file>} don't specify the file extension
% don't specify bibliographystyle
%
%%%%%%%%%%%%%%%%%%%%%%%%%%%%%%%%%%%%%%%%%%%%%%%

\bibliography{Decameter_paper_2}

%Reference citation instructions and examples:
%
% Please use ONLY \cite and \citeA for reference citations.
% \cite for parenthetical references
% ...as shown in recent studies (Simpson et al., 2019)
% \citeA for in-text citations
% ...Simpson et al. (2019) have shown...
%
%
%...as shown by \citeA{jskilby}.
%...as shown by \citeA{lewin76}, \citeA{carson86}, \citeA{bartoldy02}, and \citeA{rinaldi03}.
%...has been shown \cite{jskilbye}.
%...has been shown \cite{lewin76,carson86,bartoldy02,rinaldi03}.
%... \cite <i.e.>[]{lewin76,carson86,bartoldy02,rinaldi03}.
%...has been shown by \cite <e.g.,>[and others]{lewin76}.
%
% apacite uses < > for prenotes and [ ] for postnotes
% DO NOT use other cite commands (e.g., \citet, \citep, \citeyear, \nocite, \citealp, etc.).
%

\end{document}